\documentclass[preprint,authoryear]{elsarticle}

\usepackage{graphicx}
\usepackage{array}

\usepackage{subfigure}
\usepackage{psfrag}
\usepackage{amsmath}
\usepackage{amssymb}
\usepackage{enumerate}
\usepackage{natbib}

\def\bea{\begin{eqnarray}}
\def\eea{\end{eqnarray}}
\newcommand{\beq}{\begin{equation}}
\newcommand{\eeq}{\end{equation}}
\newcommand{\beast}{\begin{equation*}}
\newcommand{\eeast}{\end{equation*}}

\newcommand{\ie}{i.e. }
\newcommand{\eg}{e.g. }

\newcommand{\dx}{{\rm d}x}
\newcommand{\mean}{\mathrm{mean}}

\newcommand{\lambdav}{\vec \lambda}
\newcommand{\betav}{\vec \beta}
\newcommand{\capN}{\textsc n}
\newcommand{\nDR}{n_{\textsc{dr}}}
\newcommand{\pBJ}{p_{\textsc{bj}}}
\def\A{\mathcal{A}}

\def\mHz{\textrm{mHz}}
\def\Hz{\textrm{Hz}}
\def\yr{\textrm{yr}}

\begin{document}

\title
{Delayed Rejection schemes for efficient Markov chain Monte Carlo sampling of multimodal distributions}


\author[uib]{M.~Trias \corref{cor1}} \ead{miquel.trias@uib.es}
\author[bham]{A.~Vecchio} \ead{av@star.sr.bham.ac.uk}
\author[bham]{J.~Veitch} \ead{jveitch@star.sr.bham.ac.uk}

\cortext[cor1]{Corresponding author. Phone number: +34 971 172538. Fax: +34 971 173426}

\address[uib]{Departament de F\'{i}sica, Universitat de les Illes Balears, Cra. Valldemossa Km. 7.5, E-07122, Palma de Mallorca, Spain}
\address[bham]{School of Physics and Astronomy, University of Birmingham, Edgbaston, Birmingham B15 2TT, UK}


\begin{abstract}

A number of problems in a variety of fields are characterised by target distributions with a multimodal structure in which the presence of several isolated local maxima dramatically reduces the efficiency of Markov chain Monte Carlo sampling algorithms. Several solutions, such as simulated tempering or the use of parallel chains, have been proposed to facilitate the exploration of the relevant parameter space. They provide effective strategies in the cases in which the dimension of the parameter space is small and/or the computational costs are not a limiting factor. These approaches fail however in the case of high-dimensional spaces where the multimodal structure is induced by degeneracies between regions of the parameter space. In this paper we present a fully Markovian way to efficiently sample this kind of distribution based on the general Delayed Rejection scheme with an arbitrary number of steps, and provide details for an efficient numerical implementation of the algorithm.

\end{abstract}

\begin{keyword}
Degeneracies in the parameter space \sep Delayed rejection schemes \sep Efficient Markov chain Monte Carlo sampling \sep Multimodal distributions \sep Numerical implementation \sep Secondary maxima
\end{keyword}


\maketitle

\section{Introduction}
\label{sec:intro}

Bayesian inference has been gaining considerable momentum in recent years with the introduction of a number of numerical techniques to compute (marginalised) posterior density functions and marginalised likelihood (or evidence) that are at the heart of the implementation challenge of this approach. Markov chain Monte Carlo (MCMC) methods \citep{Hastings:1970, Gilks:1995, Gamerman:1997} are in particular popular and have shown the potential to tackle successfully these problems in a wide variety of fields. These methods become increasingly computationally expensive as the dimensionality and complexity of the target distribution grows, and much effort is being devoted both at the theoretical and implementation level to improve the efficiency and robustness of these approaches.

Metropolis-Hastings algorithms are employed in the overwhelming majority of the practical implementations of MCMC methods. In Metropolis-Hastings updates, a new state (or value of the parameters describing the problem at hand) is drawn from a proposal distribution and then, either accepted or rejected, depending on the value of the target distribution, with probability given by Equation~(\ref{e:alpha}). If the proposed state is rejected, the chain remains in the current state and the previous steps are repeated. Reducing the number of rejected proposals while still exploring the structure of the target distribution represents one of the most important goals in every MCMC application. By doing this, one improves the MCMC algorithm in the \cite{Peskun:1973} sense. 

\begin{figure}
\begin{tabular}{ccc}
\includegraphics[width=0.32\textwidth]{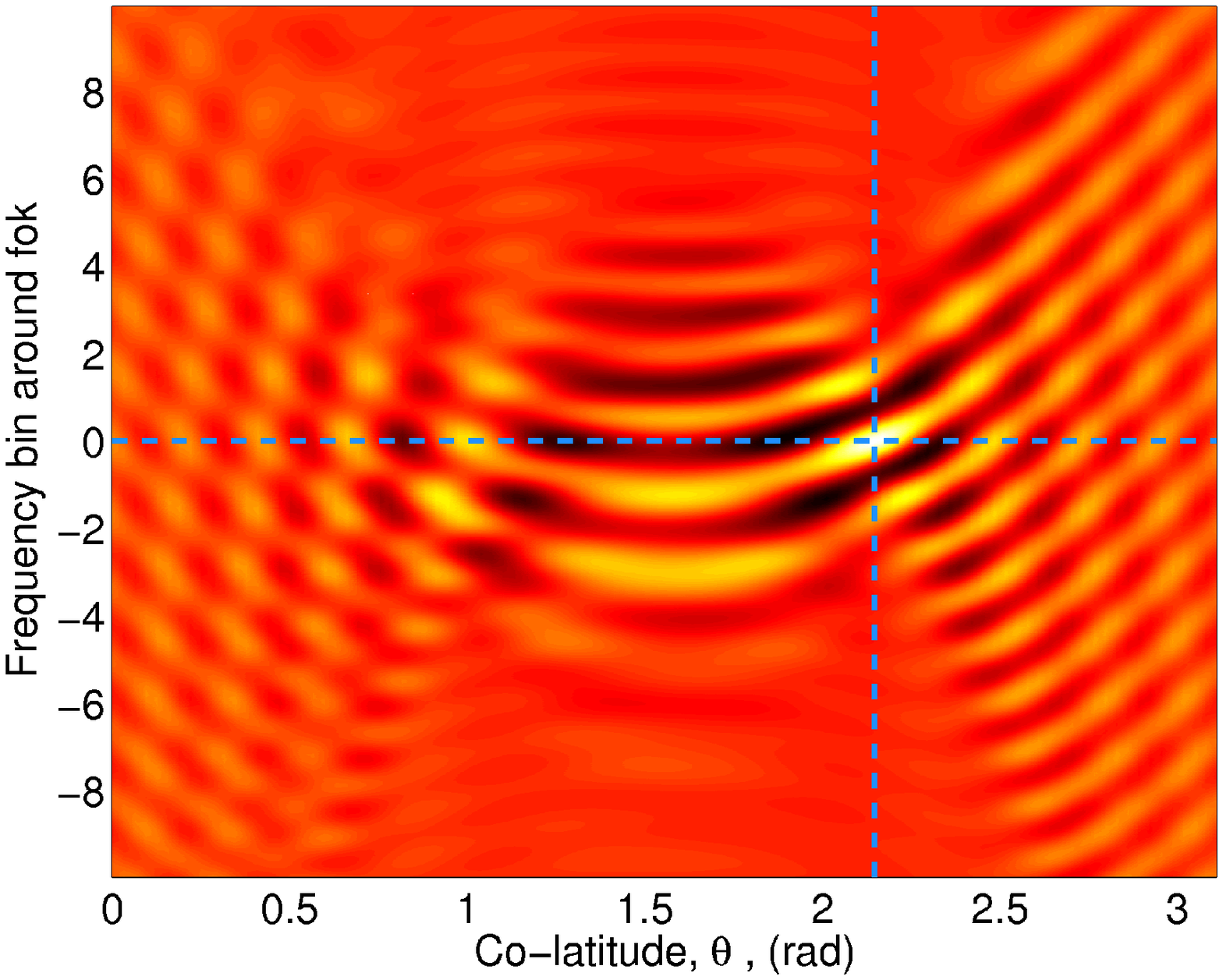} \hspace{-0.35cm} &
\includegraphics[width=0.32\textwidth]{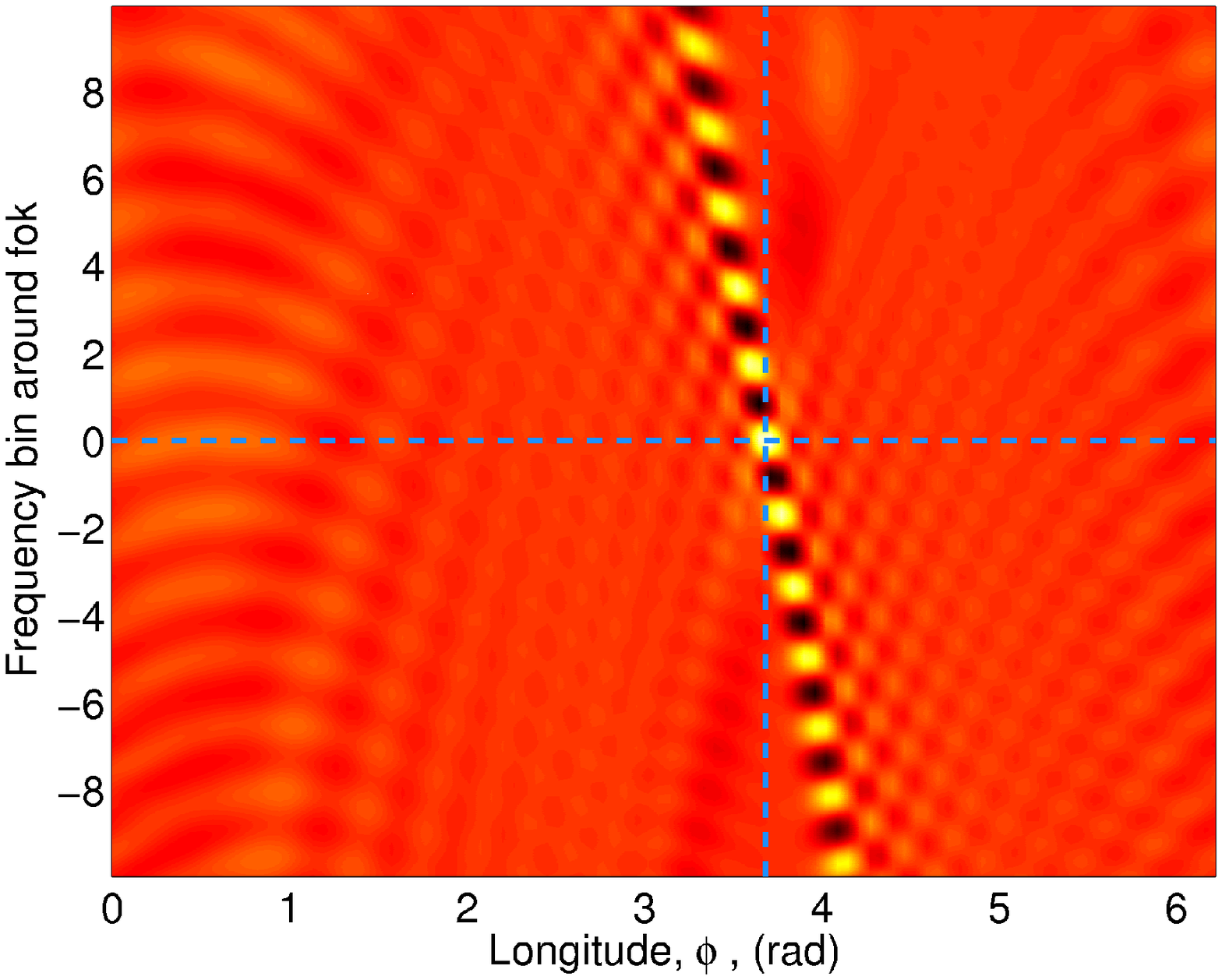} \hspace{-0.35cm} &
\includegraphics[width=0.32\textwidth]{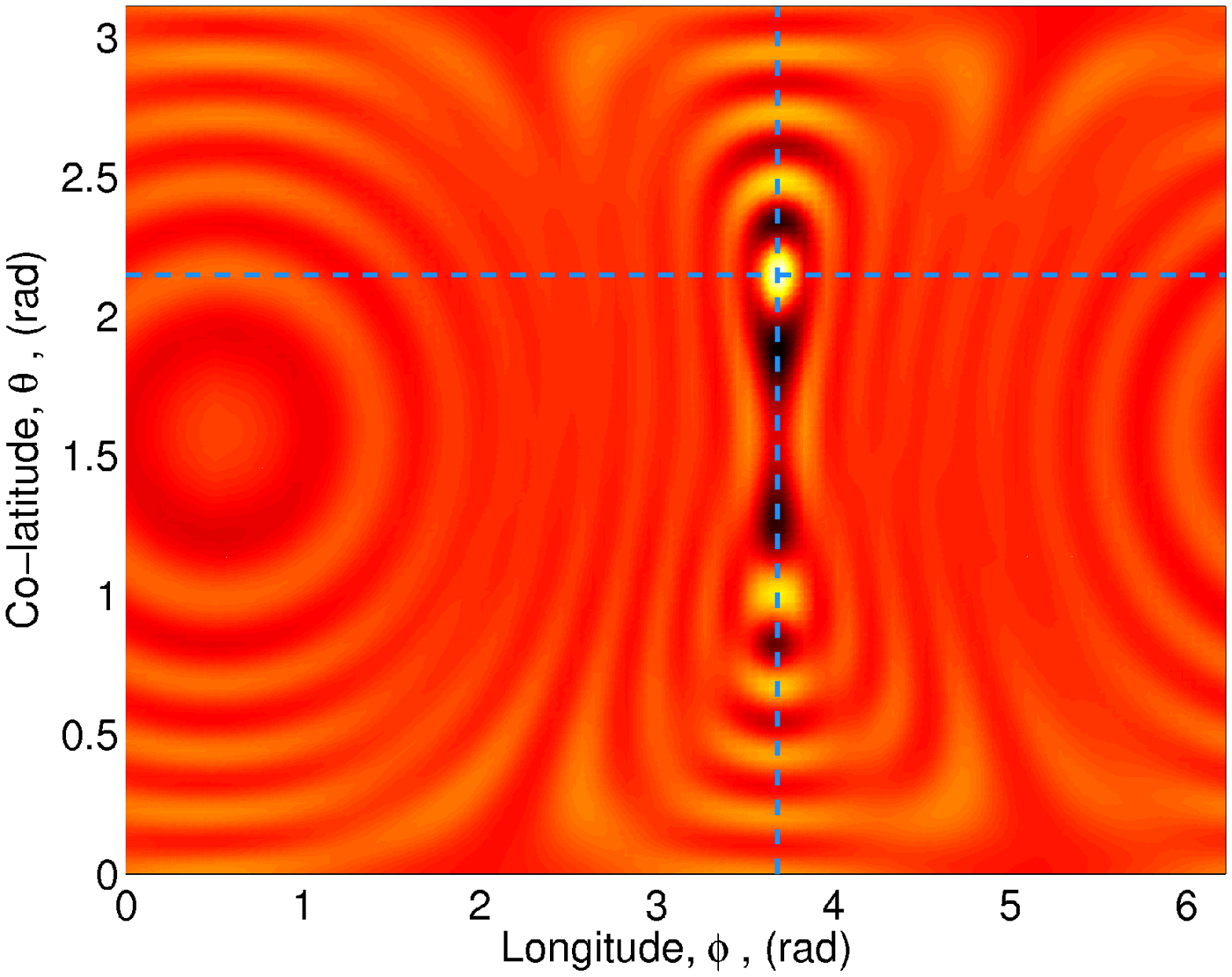}
\end{tabular}
\caption{Example of a multimodal target distribution where a normal MCMC algorithm becomes very inefficient. These figures show the colour plots of the likelihood function from a simulated gravitational wave signal from galactic white-dwarf binary systems observed with the Laser Interferometer Space Antenna taking ``slices'' through the parameter space in the three parameters that are most responsible for the structure of the likelihood function: \emph{(left)} frequency--colatitude, \emph{(center)} frequency--longitude and \emph{(right)} colatitude--longitude. The signal's frequency is given in units of the characteristic modulation frequency for this problem ($f_m = 1~\yr^{-1}$) \citep{Cutler:1998} and the two sky location angles are referenced to the ecliptic plane. The dashed-light lines represent the actual value of the parameters.
}
\label{Fig.likelihood}
\end{figure}

In a variety of practical applications, an additional complication is encountered, which is that the target distribution presents a multimodal structure, often with secondary maxima well separated from the mode (see Figure~\ref{Fig.likelihood}). In such situations, maintaining a high acceptance ratio while sampling the full structure of the target distribution becomes very difficult, leading to a very small efficiency of the algorithm. Several solutions have been proposed to increase the mixing and exploration ability of the chains, such as using parallel chains exploring the parameter space that, at a certain point, can be swapped (\eg Metropolis-coupled MCMC algorithms, see \cite{Geyer:1991}) or other methods like simulated annealing \citep{Kirkpatrick:1983} or simulated tempering \citep{Marinari:1992, Geyer:1995}, that consist in adding a `temperature' factor to the target distribution in order to make it smoother at the beginning and therefore to promote the movement of the chain. Although these techniques work fairly well in low dimensional problems, it has been shown \citep{Cornish:2008} that they are not the best solution in cases where the dimensionality of the parameter space is large and the target distribution presents a multimodal structure in several dimensions at the same time, \ie the parameter space shows very localised `islands' in which the function reaches high values, as it can be seen in Figure~\ref{Fig.likelihood}. In fact, evolving several parallel chains is computationally inefficient when the multimodal distribution only appears in some of the parameters. Simulated tempering techniques can help to improve the mixing of Markov chains, but in a high number of dimension this can hinder convergence, as shown by \cite{Cornish:2008}.

A problem that exhibits such multimodal distributions is the analysis of data from the Laser Interferometer Space Antenna \citep{Bender:1998}, which provides the motivation for our work. Figure~\ref{Fig.likelihood} shows slices through an 8-dimensional parameter space, in the plane of the three parameters most responsible for the complicated structure of the target distribution. In order to sample this function, we desire an algorithm which is both fully Markovian, and efficient in exploring the parameter space. Further details of the specific application are given in Section~\ref{sec:application}\,.

Here we present a fully Markovian method to efficiently explore the parameter space of this kind of problems, assuming that we have \emph{some} prior knowledge about the typical scales of the structures of the target distribution in the relevant parameters. In general, a normal MCMC algorithm will be able to widely explore the parameter space until `finding' one of the modes of the target distribution that will be sampled properly. The inefficiency comes from the fact that having a structure of the target distribution made up by different isolated modes (`islands') in several dimensions, it becomes very unlikely to `jump' between the different modes with normal proposals. By having a rough idea of the typical distances in the parameter space between two local maxima, one can attempt a big jump in the parameter space trying to fall in the neighbourhood of another mode (probably far away from the maximum), followed by multiple small steps exploring the new region of the parameter space (see Figure~\ref{Fig.TwoMountains}). If, during this exploration the chain finds a higher maximum, the chain would continue evolution from the new mode, but if not it would remain in the original mode. To implement this idea it is necessary to find some algorithm that allows delaying the rejection, but preserves the Markovian properties of the resulting chain.

A powerful method to overcome the rejection of proposed states and therefore to increase the efficiency of Metropolis-Hastings MCMC methods was proposed by \cite{TierneyMira:1999, Mira:2001} for fixed dimension problems and then extended by \cite{GreenMira:2001} to transdimensional problems, and goes under the name of \emph{Delayed Rejection} (DR). In short, with this method, when a new state is proposed and rejected, one can actually make a second proposal, in principle based on a different distribution which may depend on the rejected value, and make a decision (accept or reject) using a suitable probability. If the second value is rejected, one can then consider a third attempt and so on. Through this approach, one can achieve efficiency gains, while preserving the Markovian properties of the chain. In the recent years the Delayed Rejection method has been successfully applied to a number of problems of very different areas, such as particle filters \citep{Robert}, volatility models \citep{Raggi:2005}, gravitational wave searches \citep{Umstatter:2004} or object recognition \citep{Harkness:2000} and also it has been used in combination with other techniques to create high efficient methods, \eg by \cite{Al-Awadhi:2004, Haario:2006}. In all these applications of the DR algorithm, either the number of steps was very small ($2$--$3$) or the assumption of symmetrical proposals was made, simplifying the general expression for the acceptance probability of the DR chain from Eq.~(\ref{eq:mastereq}) here to Eq.~(8) of \cite{Mira:2001}. The problem we are dealing with in this paper requires the application of the full general scheme of the DR algorithm with an arbitrary number of stages. Since this is the first time to our knowledge that this has been done, in Section~\ref{sec:implementation_issues} we explain all the non-trivial implementation issues.

Moreover, MCMC methods in general and the DR approach in particular, do not provide a well defined algorithm but just the general scheme to tackle the problem. In general, actual algorithms that are efficient and robust are problem specific and require a number of complex decisions, in particular regarding choices of proposal distributions. As the number of dimensions grows and the structure of the target distribution increases, these decisions are highly non-trivial. In this paper we provide a detailed analysis of the Delayed Rejection approach and provide a set of simple and general rules for the implementation of this scheme in order to efficiently explore complicated multimodal target distributions.

The paper is organised as follows: in Section~\ref{sec:DR_theory} we review the basic concepts behind Delayed Rejection 
and demonstrate the advantage of this algorithm over a standard Markov chain exploration by comparing the variances of estimates made from the chains; in Section~\ref{sec:golden_rules} we provide general key rules for the implementation of the algorithm and useful results for the selection of the proposal distributions, and in particular, Section~\ref{subsec:parameters} summarises the actual procedure for the choice of parameters of the algorithm. In Section~\ref{sec:implementation_issues} we focus on the implementation issues related to the application of the general DR algorithm with an arbitrary number of stages; Section~\ref{sec:application} provides an application of the algorithm to the case of an 8-dimension parameter problem related to the search for gravitational waves from white dwarf binary systems in (mock) data of the Laser Interferometer Space Antenna and Section~\ref{sec:conclusion} contains our conclusions and pointers to future work. Appendix~\ref{ap:Analytical_results} provides some details about the derivation of the analytical results presented in the main body of the paper.


\section{The Delayed Rejection approach}
\label{sec:DR_theory}

The Metropolis-Hastings algorithm is widely used in applications of Markov chain Monte Carlo methods to sample a \emph{target distribution} $\pi(x)$.  Given a state of the Markov chain $\lambdav$, one proposes a new state $\lambdav'$  drawn from a \emph{proposal distribution} (or transition kernel) $q(\lambdav, \lambdav')$, the probability of $\lambdav'$ given $\lambdav$. This new state is accepted with probability
\beq\label{e:alpha}
	\alpha(\lambdav , \lambdav') = 1 \wedge \left\lbrace \frac{\pi(\lambdav') \, q(\lambdav' , \lambdav)}
							{\pi(\lambdav) \, q(\lambdav , \lambdav') }\right\rbrace \; ,
\eeq
and the chain remains at $\lambdav$ with probability $1 - \alpha(\lambdav , \lambdav')$. In the previous equation, the notation $a \wedge b$ (for any real numbers $a$ and $b$) stands for the minimum between $a$ and $b$, and as a consequence the acceptance probability is limited to unity in the case where the ratio is $>1$.
The convergence to the target distribution $\pi$ is guaranteed through the acceptance probability, but the convergence rate or efficiency of the chain is highly dependent on the choice of the transition kernel $q$. Its choice is one of the most critical stages in actually implementing a Markov chain Monte Carlo algorithm. An effective way to improve the efficiency of the algorithm has been proposed in the form of Delayed Rejection by \cite{TierneyMira:1999, Mira:2001, GreenMira:2001} which we now briefly summarise mainly to establish notation for the case of a problem with a fixed number of dimensions. 

\subsection{Delayed Rejection}
\label{subsec:general_DR}

Suppose the current state of the chain is $\lambdav$. A candidate, $\betav_1$, is generated from a proposal distribution $q_1(\lambdav , \betav_1)$ and accepted with probability
\beq
	\alpha_1(\lambdav , \betav_1) = 1 \wedge \left\lbrace \frac{\pi(\betav_1) \, q_1(\betav_1 , \lambdav)}
	 																					     {\pi(\lambdav) \, q_1(\lambdav , \betav_1) }\right\rbrace \; ,
\eeq
as in a standard Metropolis-Hastings algorithm, see Equation~(\ref{e:alpha}). If $\betav_1$ is not accepted, instead of rejecting it, one can use this information and propose a new state $\betav_2$ from a (in principle different) proposal distribution $q_2(\lambdav , \betav_1 , \betav_2)$ which may use information about the rejected state $\betav_1$. In order to maintain the same stationary distribution, the acceptance probability of this new candidate may be computed as
\beq
	\alpha_2(\lambdav, \betav_1 , \betav_2) = 1 \wedge \left\lbrace
	\frac{\pi(\betav_2) \, q_1(\betav_2 , \betav_1) \, q_2(\betav_2 , \betav_1 , \lambdav) \left[ (1-\alpha_1(\betav_2 , \betav_1)\right] }
			{\pi(\lambdav) \, q_1(\lambdav , \betav_1) \, q_2(\lambdav , \betav_1 , \betav_2) \left[ (1-\alpha_1(\lambdav , \betav_1)\right] }
	\right\rbrace \; ,
\eeq
which is derived by imposing that the acceptance probability must preserve the detailed balance separately at each stage.

This procedure can be applied for an arbitrary number of stages: the result is a generalisation of Markov chain Monte Carlo methods that is called Delayed Rejection (DR). The generic $i-$th stage of the chain is as follows: if the state $\betav_{i-1}$ is proposed and rejected, one can propose a new candidate $\betav_i$ from the proposal $q_i(\lambdav , \betav_1 , \ldots , \betav_i )$ and accept it with probability
\bea \label{eq:mastereq}
\alpha_i (\lambdav , \betav_1, \ldots , \betav_i) & = &
1 \wedge \left\lbrace \frac{\pi(\betav_i)~q_1(\betav_i,\betav_{i-1})~q_2(\betav_i,\betav_{i-1},\betav_{i-2})~\ldots ~
q_i(\betav_i,\betav_{i-1}, \ldots , \lambdav )}
{\pi(\lambdav )~q_1(\lambdav ,\betav_1)~q_2(\lambdav ,\betav_1,\betav_2)~\ldots ~q_i(\lambdav ,\betav_1, \ldots , \betav_i)} \right.
\nonumber\\
& & \hspace{-1.5cm} \left. \frac{\left[ 1-\alpha_1(\betav_i,\betav_{i-1}) \right] \left[ 1-\alpha_2(\betav_i,\betav_{i-1},\betav_{i-2}) \right] 
\ldots \left[ 1-\alpha_{i-1}(\betav_i, \ldots \betav_1) \right] }
{\left[ 1-\alpha_1(\lambdav ,\betav_1) \right] \left[ 1-\alpha_2(\lambdav ,\betav_1,\betav_2) \right] \cdots 
\left[ 1-\alpha_{i-1}(\lambdav , \ldots \betav_{i-1}) \right] } \right\rbrace \; .
\eea
Notice that with the notation we are using, $q_i(\lambdav ,\betav_1, \ldots , \betav_{i-1} , \betav_i)$ represents the proposal probability of $\betav_i$ given $\{ \lambdav , \betav_1 , \ldots , \betav_{i-1} \}$ where the order of the parameters does matter. Sometimes, in the literature this also may be written as $q_i(\betav_i | \lambdav , \betav_1 , \ldots , \betav_{i-1})$.

Due to the way in which the DR algorithm has been designed, \ie by imposing detailed balance at each stage and deriving the acceptance probability that preserves it, the resulting chain is a reversible Markov chain with invariant distribution $\pi$. Thus, the average along a sample path of the chain of a function $f$, is an asymptotically unbiased estimate of $\int f(x) \pi(x) \dx$.

The benefit of this method is that it increases the efficiency of sampling by delaying a possible rejection, see Sec.~\ref{subsec:benefits_DR}; furthermore, it allows us to propose a new candidate that can be drawn using the information of the past elements. These features make the Delayed Rejection algorithm very useful in a number of applications; 
in particular we shall use it to increase the efficiency in computing marginalised posterior distributions generated by integrating multimodal likelihood functions.

An essential point of the DR scheme is that Eq.~(\ref{eq:mastereq}) has been derived by imposing that the backward path, from $\betav_i$ to $\lambdav$, follows the forward path, from $\lambdav$ to $\betav_i$, in reverse order. As a consequence, this property forces one to choose proposal distributions that preserve, in some way, the reversibility of the chain. As we shall show, this choice is complicated in any practical application; despite this fact, at each stage we will have the freedom to propose a new stage with a different proposal with respect to the previous one and to make use of all the past information. In the next section, Sec.~\ref{sec:golden_rules}, we discuss the details of this key point and provide a general method to actually obtain a Markov chain with Delayed Rejection, regardless of the specific target distribution.

\subsection{Efficiency increase with Delayed Rejection}
\label{subsec:benefits_DR}

When a new state is proposed and rejected, instead of retaining the previous state and duplicating an element of the Markov chain, the DR algorithm allows us to make a new proposal and consider it for acceptance. Naturally, this increases the transition probability of the elements of the chain within the different parts of the parameter space, improving the mixing of the chain and leading to a reduction of the asymptotic variance of any estimate \citep{Peskun:1973}.

A more quantitative statement about the benefits of using the Delayed Rejection algorithm when a transition with low acceptance probability is attempted, can be made by computing the variance of an estimate, $\bar\lambda$, using the autocorrelation function of the Markov chain \citep{Sokal:1989, Gilks:1995, Gelman:1996},
\beq \label{eq:var_vs_tau}
\textrm{var} (\bar{\lambda}) \simeq \frac{2 ~ \tau_{int} ~ C_{\lambda \lambda} (0)}{N} \; ,
\eeq
where $N$ is the number of elements of the Markov chain, $C_{\lambda \lambda} (t)$ is the autocorrelation function of the parameter $\lambda$ at lag $t$ and
\beq \label{eq:tau}
\tau_{int} = \frac{1}{2} + \sum_{n=1}^{N} \frac{C_{\lambda \lambda} (n)}{C_{\lambda \lambda} (0)} \equiv \frac{1}{2} + \sum_{n=1}^{N} \rho_{\lambda \lambda} (n) \; .
\eeq
From a theoretical point of view, we can demonstrate that the variance of an estimate made from a chain using Delayed Rejection is always smaller than that produced with a standard Markov chain.

\begin{figure}
	\centering
	\includegraphics[width=0.85\textwidth]{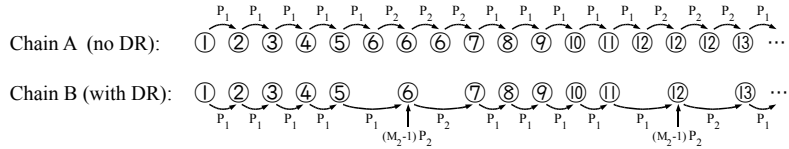}
	\caption{Example of the Markov chains that would be generated in the theoretical case considered in Sec.~\ref{subsec:benefits_DR}: every $M_1$ elements drawn from a certain proposal, $P_1$, $M_2$ transitions from a second proposal, $P_2$, are attempted but accepted with lower probability. The Delayed Rejection algorithm attempts the same number ($M_2$) of transitions drawn from $P_2$ as the standard MCMC, but it only stores the one that is accepted. Different circled numbers represent possible different elements of the chain, whereas the repeated ones explicitly denote the rejected transitions drawn from $P_2$. In this example we are considering $M_1 = 5$ and $M_2 = 3$.}
\label{f:chains4eff}
\end{figure}

Let us consider a Markov chain generated from two different proposals, $P_1$ and $P_2$; one of them, say $P_2$, has a lower acceptance probability. To establish notation, we assume that after $M_1$ elements drawn from $P_1$, $M_2$ transitions from $P_2$ are attempted and all of them are rejected except for the last one (see Fig.~\ref{f:chains4eff}). This will generate a chain ($A$) with $M_2$ completely correlated (repeated) elements every $M_1+M_2$ iterations, whose efficiency could be improved by grouping all these $M_2$ elements into a single one using the DR algorithm (chain $B$ in Fig.~\ref{f:chains4eff}). Removing repeated elements means reducing the autocorrelation of the chain, $\tau_{int}$, or equivalently increasing the transition probability and, therefore, getting lower asymptotic variances of any estimate \citep{Peskun:1973}; but it also means reducing the number of elements, $N$, of the chain for the same computational cost. So, qualitatively it is not clear which effect will dominate in Eq.~(\ref{eq:var_vs_tau}), but one can explicitly derive the expression that relates the variances of the same estimate using chains $A$ and $B$. With our notation, this is:
\bea \label{e:benefitsDR} \small
\frac{\textrm{var}_A (\bar{\lambda}) - \textrm{var}_B (\bar{\lambda})} {\textrm{var}_B (\bar{\lambda})} & = & \small \left( \frac{M_2-1}{M_1+M_2} \right) ^2 \left[ (M_1+1) ~ \frac{\frac{1}{2} + \sum\limits_{n=1}^{N/(M_1+1)} \rho_{\lambda \lambda} \left( (M_1+1) n \right) } {\frac{1}{2} + \sum\limits_{n=1}^{N} \rho_{\lambda \lambda} (n) } - 1\right] = \nonumber \\
  & & \hspace{-1.5cm} = \left( \frac{M_2-1}{M_1+M_2} \right) ^2 \left[ (M_1+1) \coth\left( \tfrac{M_1+1}{2 ~ \tau_{exp}} \right) \tanh\left( \tfrac{1}{2 ~ \tau_{exp}} \right) -1   \right] \geq 0 \; ,
\eea
where we have used the definition of the exponential autocorrelation time, $\rho_{\lambda \lambda} (n) = \exp(-|n|/\tau_{exp})$. Equation~(\ref{e:benefitsDR}) demonstrates that, for a fixed computational time, \emph{we always obtain a benefit from using the Delayed Rejection algorithm in terms of reducing the variance of any estimate}; furthermore, this equation provides a quantitative result for this statement.


\section{Implementing Delayed Rejection schemes: Key rules}
\label{sec:golden_rules}

The Delayed Rejection method presented in a formal way in the previous section, guarantees the stationarity and Markovian property of the resulting chain; concurrently, it provides ample freedom (i) to choose different proposal distributions at each stage of the DR algorithm and (ii) to use all the information from previous values of the chain. What is not guaranteed is the ability to achieve a non-negligible acceptance probability at some stage of the DR chain. In fact, we shall see that in the general case without following some key rules about the choice of proposals, the acceptance probability always tends to zero, which in turn would defeat the very reason to consider a Delayed Rejection algorithm.

At the $i$-th state of the DR chain, $\betav_i$ is generated from $q_i(\lambdav , \betav_1 , \ldots , \betav_i )$ and accepted with an acceptance probability given by Equation~(\ref{eq:mastereq}). This expression is the result of the product of three different kinds of terms: 
\begin{itemize}
\item The likelihood ratio between the new proposed element of the chain, $\betav_i$, and the first one, $\lambdav$:
\beq\label{eq:like_ratio}
\frac{\pi(\betav_i)}{\pi(\lambdav)}\,;
\eeq
\item The ratio of proposal probabilities
\beq\label{eq:prop_ratio}
 \frac{q_1(\betav_i,\betav_{i-1})~q_2(\betav_i,\betav_{i-1},\betav_{i-2})~\ldots ~
q_i(\betav_i,\betav_{i-1}, \ldots , \lambdav )}
{q_1(\lambdav ,\betav_1)~q_2(\lambdav ,\betav_1,\betav_2)~\ldots ~q_i(\lambdav ,\betav_1, \ldots , \betav_i)}\,;
\eeq
\item The ratio of complementary acceptance probabilities,
\beq\label{eq:accept_ratio}
\frac{\left[ 1-\alpha_1(\betav_i,\betav_{i-1}) \right] \left[ 1-\alpha_2(\betav_i,\betav_{i-1},\betav_{i-2}) \right] 
\ldots \left[ 1-\alpha_{i-1}(\betav_i, \ldots \betav_1) \right] }
{\left[ 1-\alpha_1(\lambdav ,\betav_1) \right] \left[ 1-\alpha_2(\lambdav ,\betav_1,\betav_2) \right] \cdots 
\left[ 1-\alpha_{i-1}(\lambdav , \ldots \betav_{i-1}) \right] }\,.
\eeq
\end{itemize}
In Section~\ref{sec:implementation_issues} we will show that the latter term is (in most of the cases) of order unity, and we will discuss how to compute it efficiently. Following the idea of a Metropolis-Hastings ratio \citep{Gilks:1995, Gamerman:1997}, we would like the acceptance probability~(\ref{eq:mastereq}) to be dominated by the likelihood ratio, Equation~(\ref{eq:like_ratio}), and therefore to have typical values for the proposals ratio~(\ref{eq:prop_ratio}) of order unity; this will only be valid if the \emph{reversibility} of the DR chain, drawn from the different proposal distributions, is preserved. Having a handle on the ratio of proposal probabilities is therefore essential for a Delayed Rejection scheme, and we will now concentrate on this specific point. 

Let us focus now on the ratio of proposal probabilities, Equation~(\ref{eq:prop_ratio}). In the denominator, we are evaluating proposal probabilities of the \emph{normal chain}, in the sense that we are computing the probability of proposing $\betav_j$ ($j<i$) given $\{ \lambdav , \betav_1, \ldots , \betav_{j-1} \}$ when $\betav_j$ was indeed generated from $q_j(\lambdav , \betav_1 , \ldots , \betav_j )$. On the other hand, the proposal probabilities at the numerator correspond to the \emph{reverse chain}, \ie we are computing what would be the probability of proposing $\betav_j$ ($j<i$) assuming that the old elements of the chain are $\{ \betav_i, \betav_{i-1}, \ldots, \betav_{j+1} \}$, when in reality $\betav_j$ was generated from $q_j(\lambdav , \betav_1 , \ldots , \betav_j )$. For this reason it can be shown (see Appendix~\ref{ap:Analytical_results}) that on average, the ratio of proposal probabilities (\ref{eq:prop_ratio}) is smaller than one; how small with respect to ${\pi(\betav_i)}/{\pi(\lambdav)}$ is at the very heart of the problem and it will dramatically affect the efficiency of the MCMC. In turn, the actual value of the term~(\ref{eq:prop_ratio})  depends on the \emph{reversibility} of the DR chain.

\begin{figure}
	\centering
    \includegraphics[width=8cm]{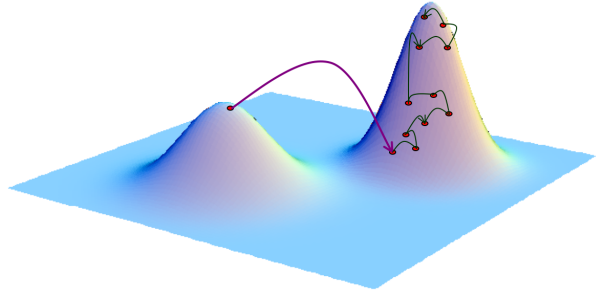}
	\caption{Graphical representation of the idea we want to implement to efficiently explore different maxima of the target distribution using the Delayed Rejection algorithm: an initial big jump (proposed by $q_a$) which may not land in the peak of the neighbourring node, followed by multiple small steps (drawn from $q_b$) in order to explore the region we landed. In this paper we show that this exploration cannot be ``up-hill'', as in a standard Metropolis-Hastings method, but blind in order to achieve reasonable acceptance probabilities.}
\label{Fig.TwoMountains}
\end{figure}

The main application we have in mind for the Delayed Rejection algorithm is a Markovian way to sample target distributions that are characterised by well separated local maxima in parameter space. In order to achieve this goal we proceed as follows: at a certain stage of a standard Metropolis-Hastings MCMC chain -- likely exploring a secondary mode of the distribution -- we start a DR chain with a big jump that attempts to reach the mode of the distribution; such transition, likely to be rejected but hopefully closer to the mode, is followed by many small ones in order to explore the new region of the parameter space (see Figure~\ref{Fig.TwoMountains} for a cartoon  exemplifying the situation):
\begin{center}
\begin{tabular}{ccm{6cm}}
$(j = 1)$   &    $q_1(\lambdav, \betav_1) \equiv q_a(\lambdav , \betav_1)$   &  \begin{center} ~~~ mainly proposing big jumps \newline (travel to another local maximum) \end{center}  \\
$(j \geq 2)$  &    $q_j(\lambdav , \betav_1 , \ldots , \betav_j) \equiv q_b(\bar{x}[\lambdav , \ldots , \betav_{j-1}] , \betav_j)$  & \begin{center} ~ mainly proposing small jumps \newline (explore the new region) \end{center}
\end{tabular}
\end{center}
Thus, it is inevitable that the resulting chain will have some degree of non-reversibility. However, this behaviour can be controlled in order to yield a non-negligible acceptance probability from Eq.~(\ref{eq:mastereq}). We will devote Section~\ref{subsec:3_gauss} to address this essential point.

From here on and for the sake of simplicity, we will assume that during the Delayed Rejection algorithm, only one parameter is being updated. All the conclusions and results contained in the following sections can be straightforwardly extended to the case of updating several independent parameters; furthermore all the guiding principles (but not the numerical results) would be unaffected for the general case of updating correlated parameters.

\subsection{Choice of proposals: Functional form}
\label{subsec:3_gauss}

The first and last terms of the product of proposal probabilities,
\beq \label{Eq.terms_AP}
\frac{q_a(\beta_i , \beta_{i-1})}{q_a(\lambda , \beta_1)} \qquad \mathrm{and} \qquad \frac{q_b(\bar{x} [\beta_i , \ldots , \beta_{1} ], \lambda )}{q_b(\bar{x} [\lambda , \ldots , \beta_{i-1}] , \beta_i )}
\eeq
are the ones which involve either the initial `big jump' proposal or the elements of the chain generated from it. In the numerator of the first fraction, we are evaluating a `big jump' proposal, $q_a$, but with two elements of the chain (mostly probably) separated by a small distance in the parameter space. The opposite case happens in the numerator of the other fraction, where two elements of the chain with a likely big separation in the parameter space are evaluated with an `small jump' proposal, $q_b$.

Having a $q_a$ proposal only contemplating big jumps and $q_b$ small jumps, would result in a tiny value for the ratio of proposal probabilities, Eq.~(\ref{eq:prop_ratio}), very difficult to compensate with the likelihood ratio in order to finally get non-negligible values for the acceptance probability in Equation~(\ref{eq:mastereq}). The solution to this problem comes through including a certain probability of doing small jumps in $q_a$ and big jump proposals in $q_b$ and therefore, all our proposals become a mixture of three Gaussian distributions (3-Gaussian, in future references), symmetrical around the central mode. 

\begin{figure}
	\includegraphics[width=\textwidth]{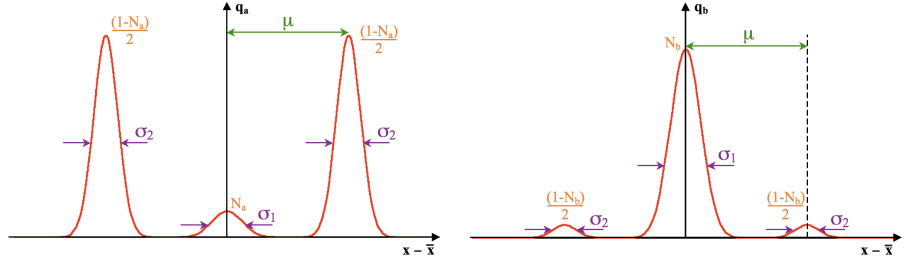}
	\caption{Proposal distributions used in the general Delayed Rejection algorithm in order to efficiently explore different maxima of the target distribution. We force the 3-Gaussian functions to be symmetrical and therefore, they can be characterised by $5$ parameters, $\{ \sigma_1 , \sigma_2 , \mu , N_a , N_b \}$, graphically represented in this figure. For each particular problem, they will be chosen to be adapted to the structure of the target distribution: $\mu$ being the typical distance between different maxima and $\sigma_i$ their typical width. The optimal values for $N_a$ and $N_b$ are found as a compromise between efficiently exploring the parameter space and having non-negligible acceptance probabilities, see Sec.~\ref{sec:golden_rules} for further discussions.}
\label{Fig.Proposals}
\end{figure}

In Figure~\ref{Fig.Proposals} we graphically represent the proposal functions that we are going to use, that can be parametrised as:
\bea \label{eq:Proposals}
q_{a,b} (\bar{x} , x) & = & \frac{1-N_{a,b}}{2} \, \frac{1}{\sqrt{2 \pi} \sigma_2} \exp\left[-\frac{(x+\mu-\bar{x})^2}{2 \sigma_2^2}\right] + \nonumber \\
                  & & N_{a,b} \, \frac{1}{\sqrt{2 \pi} \sigma_1} \exp\left[-\frac{(x-\bar{x})^2}{2 \sigma_1^2}\right] + \nonumber \\
                  & & \frac{1-N_{a,b}}{2} \, \frac{1}{\sqrt{2 \pi} \sigma_2} \exp\left[-\frac{(x-\mu-\bar{x})^2}{2 \sigma_2^2}\right]\; ,
\eea
where $\bar{x}$ may be a function of the old elements of the DR chain and $N_{a,b}$ represents the probability of proposing a value from the central Gaussian. The extreme case where $q_a$ is always proposing big jumps and $q_b$ small ones, would be given by setting $N_a = 0$ and $N_b = 1$, but a compromise must be reached between having an efficient way to explore multimodal target distributions and obtaining non-negligible acceptance probabilities.

\begin{figure}
\hspace{-1.35cm}
\begin{tabular}[t]{p{1.5cm}cccc}
 & \hspace{-0.7cm}
 \scriptsize{$\displaystyle \frac{\sigma_2}{\mu} = 2.00$} & \hspace{-0.75cm}
 \scriptsize{$\displaystyle \frac{\sigma_2}{\mu} = 0.54$} & \hspace{-0.75cm}
 \scriptsize{$\displaystyle \frac{\sigma_2}{\mu} = 0.15$} & \hspace{-0.75cm}
 \scriptsize{$\displaystyle \frac{\sigma_2}{\mu} = 0.04$}
 \\[0.3cm]
 \vspace{-1.9cm}
 \scriptsize{$\displaystyle \frac{\sigma_1}{\mu} = 2.00$}
 \vspace{1.1cm}
 & \hspace{-0.7cm}
\includegraphics[width=3.8cm]{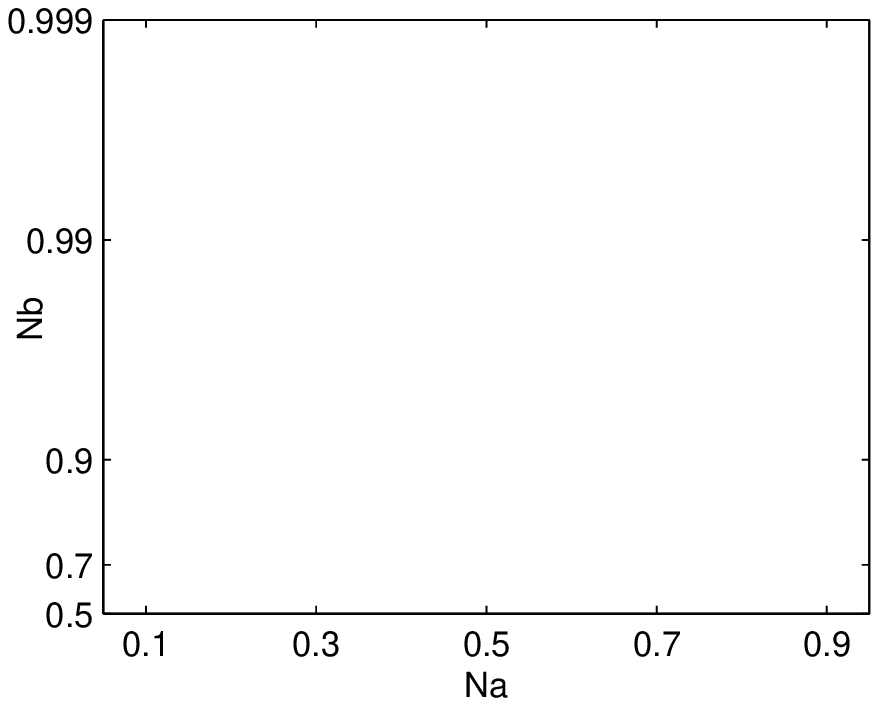}  & \hspace{-0.75cm}
\includegraphics[width=3.8cm]{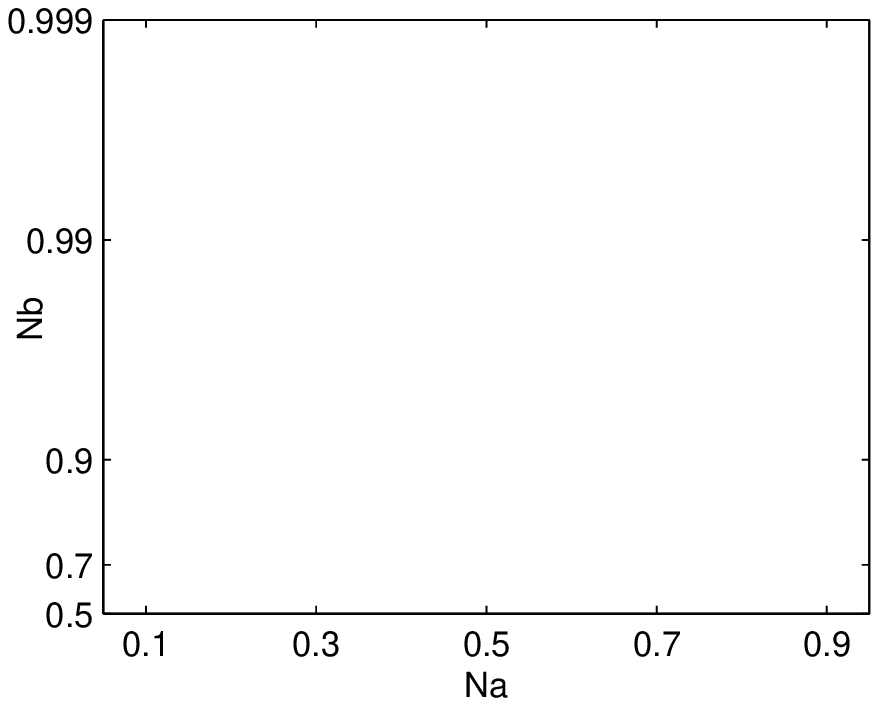}  & \hspace{-0.75cm}
\includegraphics[width=3.8cm]{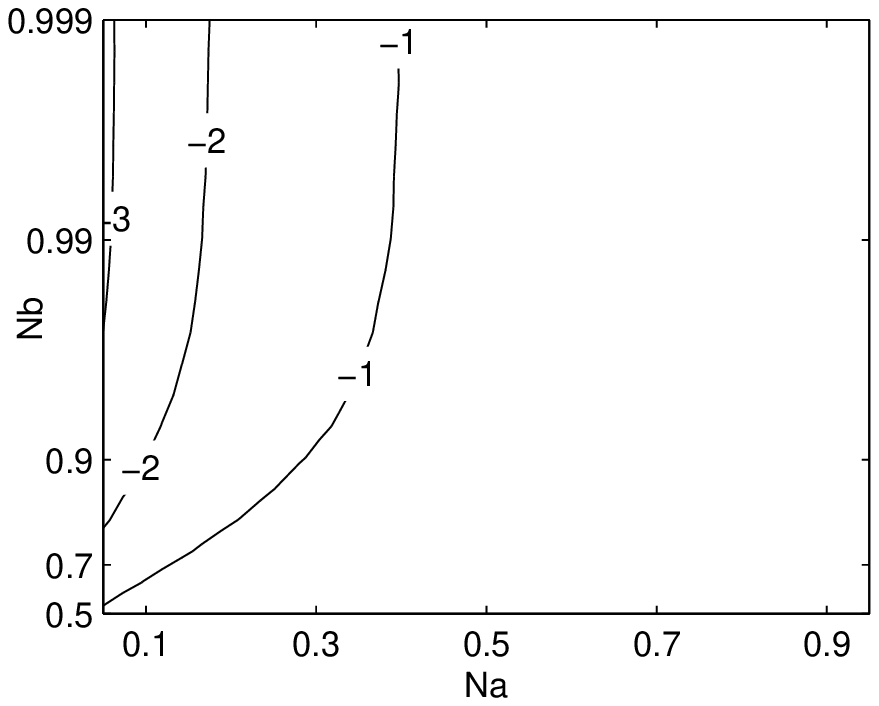}  & \hspace{-0.75cm}
\includegraphics[width=3.8cm]{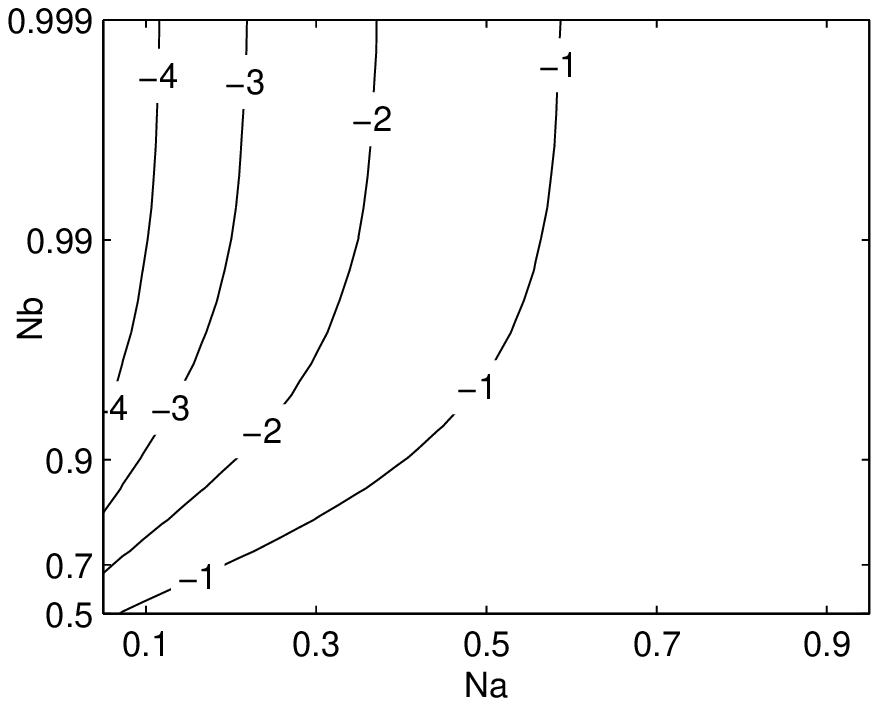}
\\
 \vspace{-1.9cm}
 \scriptsize{$\displaystyle \frac{\sigma_1}{\mu} = 0.54$} 
 \vspace{1.1cm} 
 & \hspace{-0.7cm}
\includegraphics[width=3.8cm]{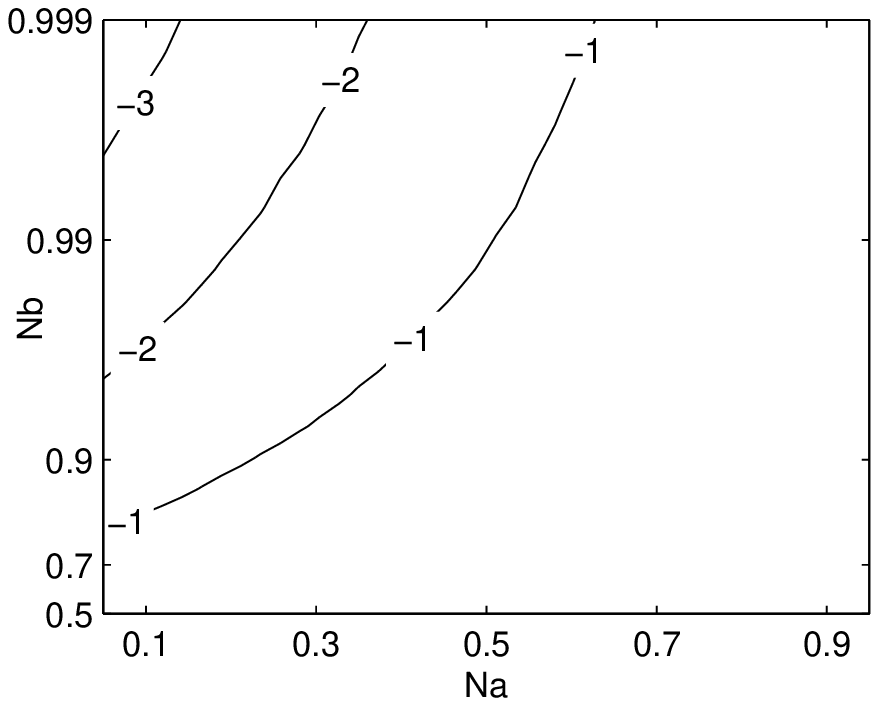}  & \hspace{-0.75cm}
\includegraphics[width=3.8cm]{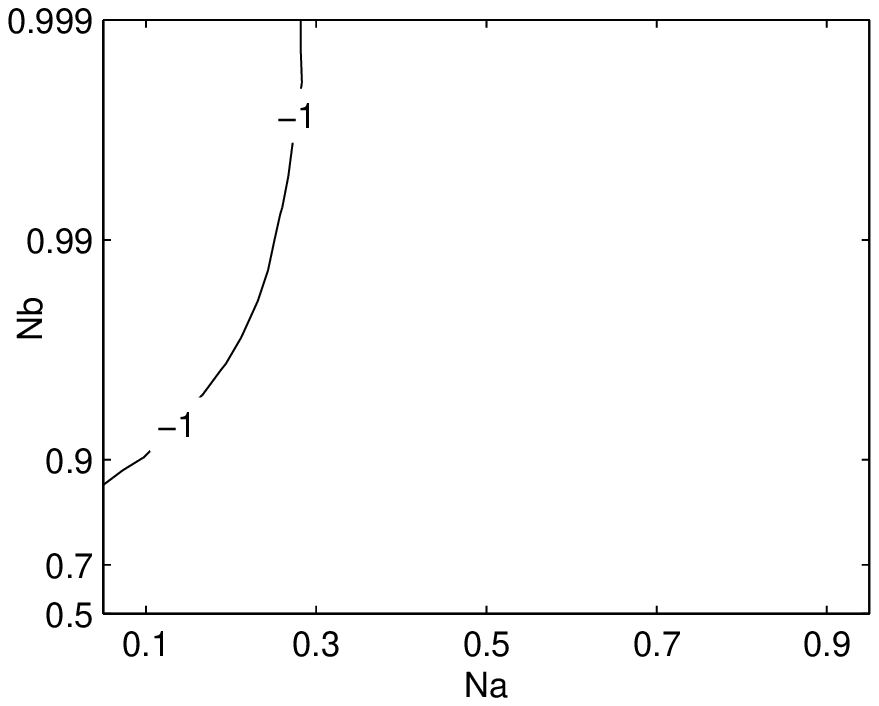}  & \hspace{-0.75cm}
\includegraphics[width=3.8cm]{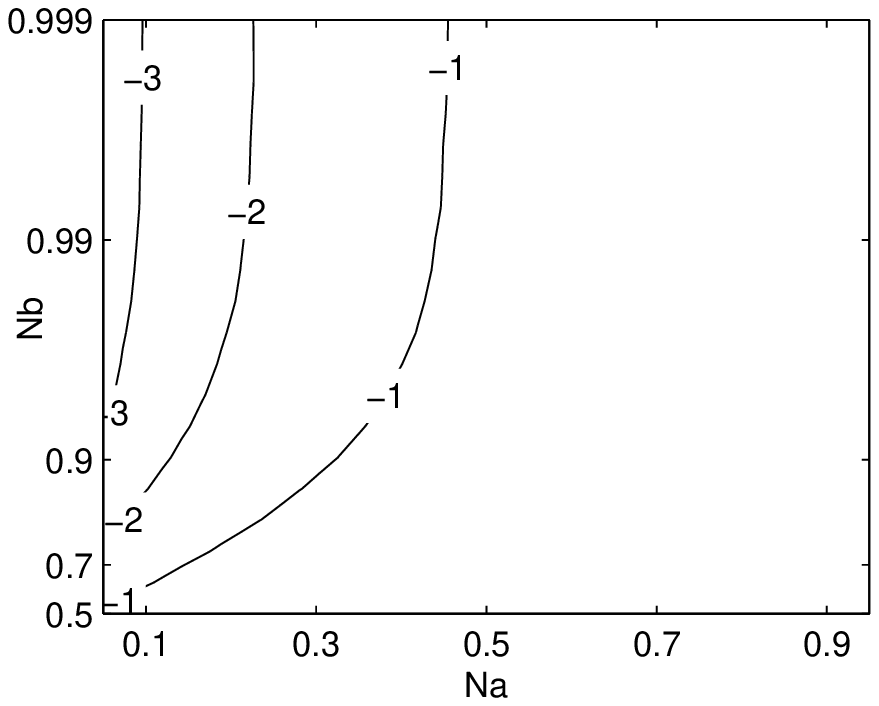}  & \hspace{-0.75cm}
\includegraphics[width=3.8cm]{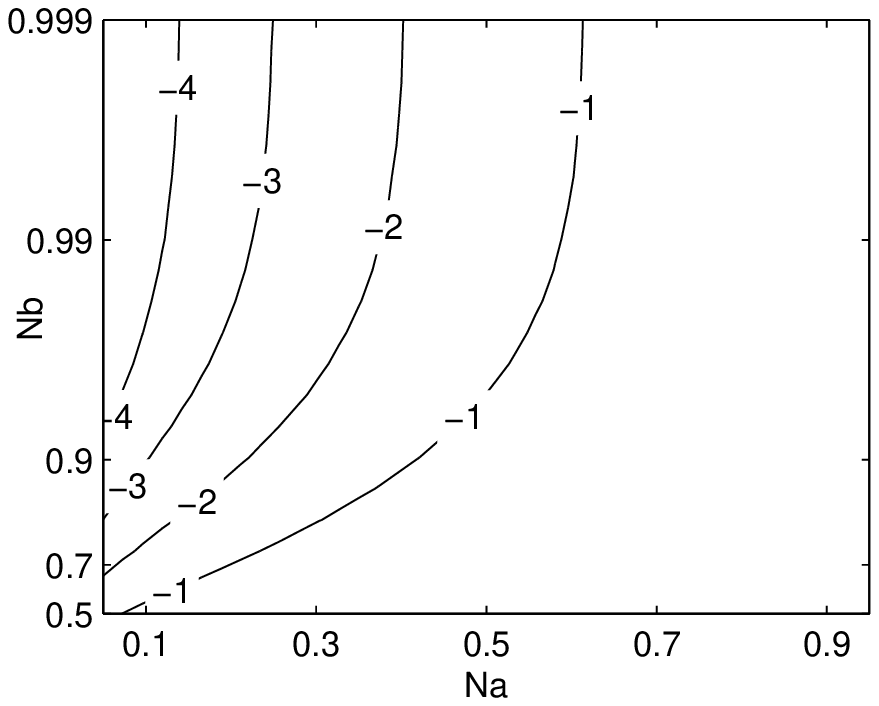}
\\
 \vspace{-1.9cm}
\scriptsize{$\displaystyle \frac{\sigma_1}{\mu} = 0.15$} 
 \vspace{1.1cm} 
 & \hspace{-0.7cm}
\includegraphics[width=3.8cm]{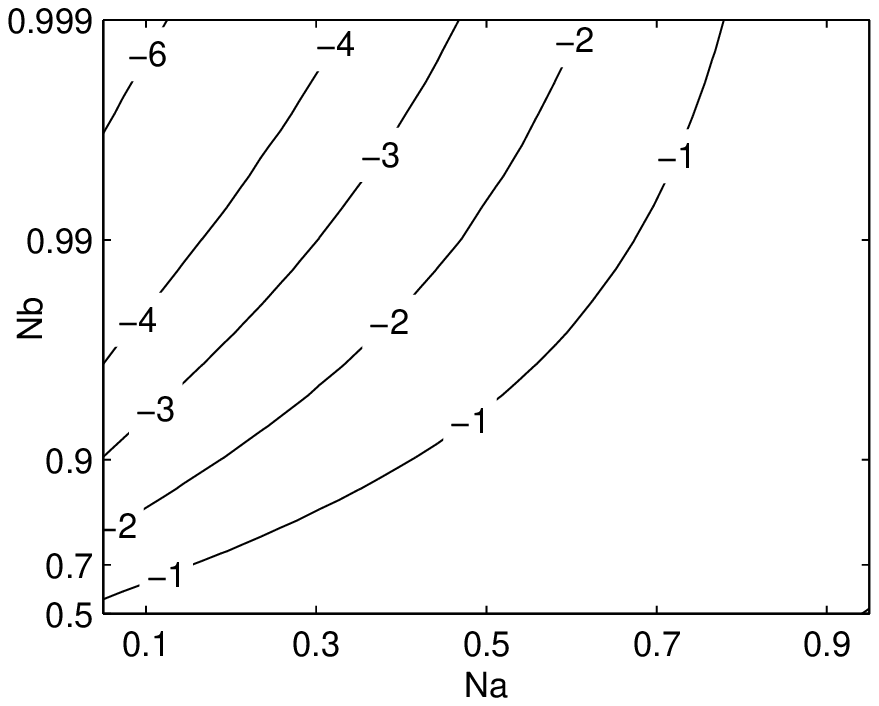}  & \hspace{-0.75cm}
\includegraphics[width=3.8cm]{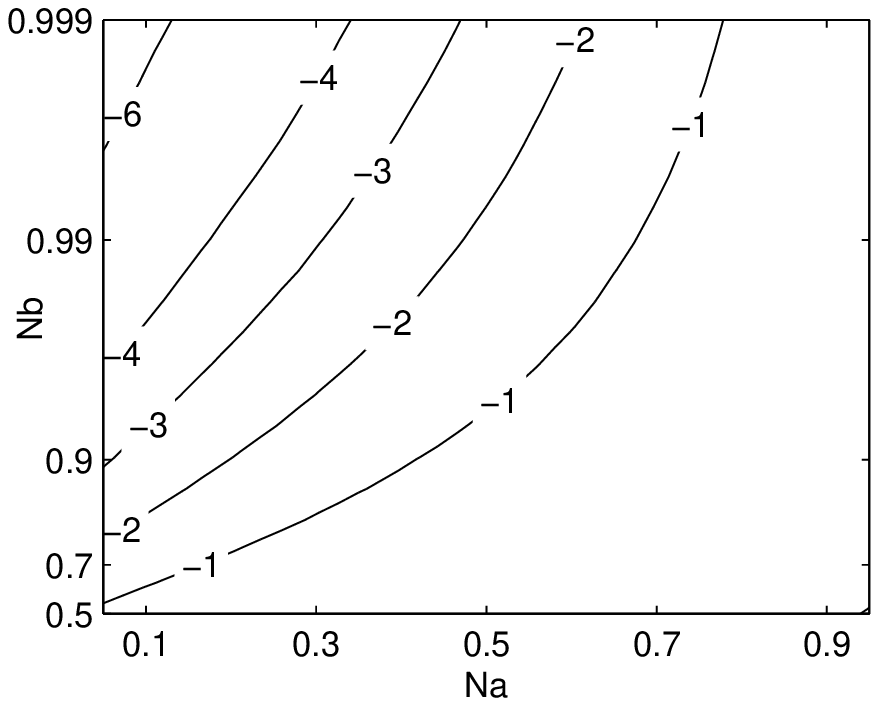}  & \hspace{-0.75cm}
\includegraphics[width=3.8cm]{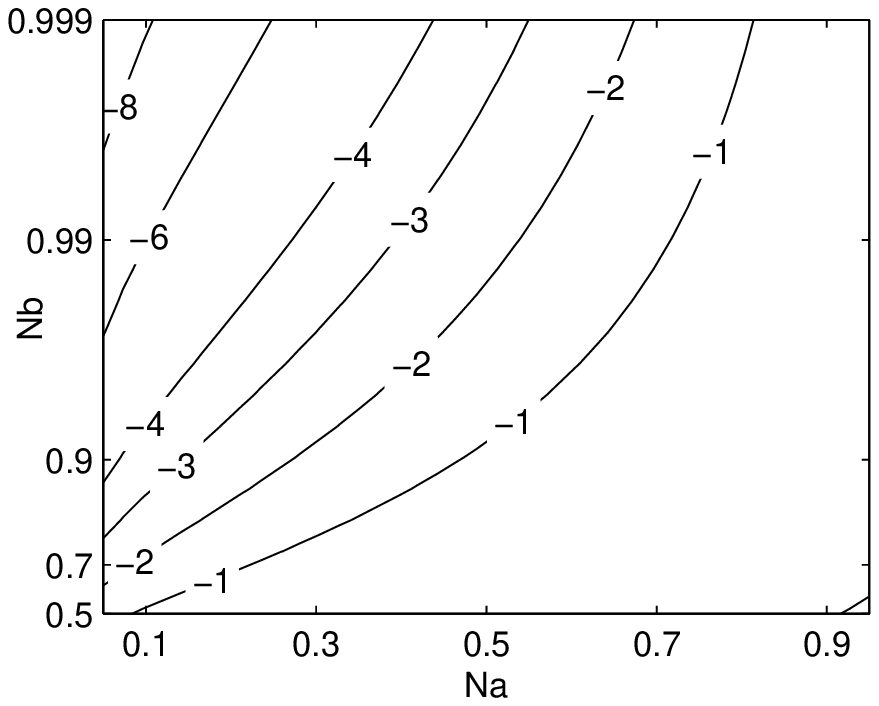}  & \hspace{-0.75cm}
\includegraphics[width=3.8cm]{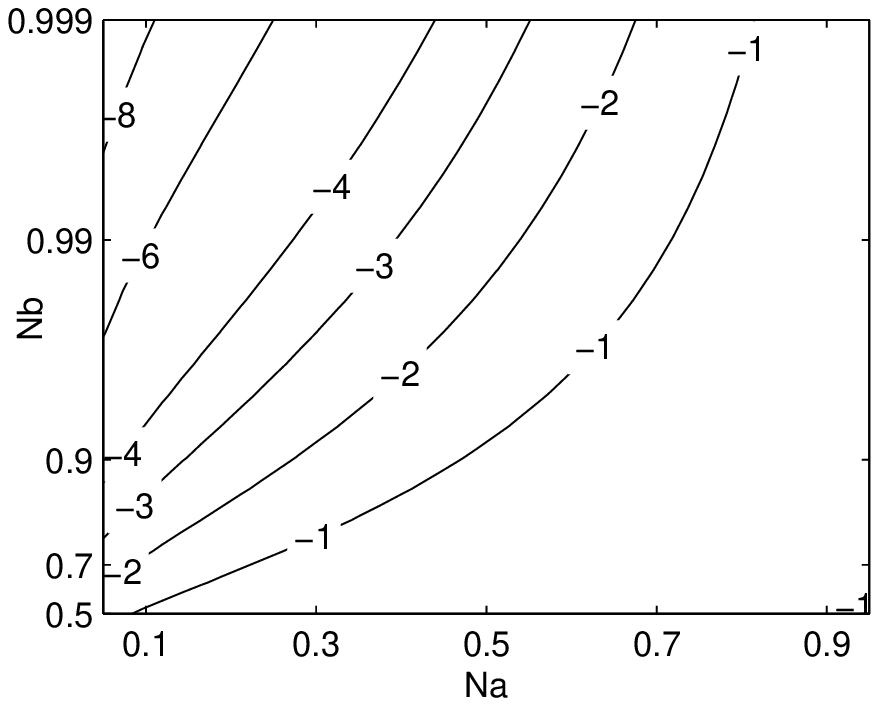}
\\
 \vspace{-1.9cm}
\scriptsize{$\displaystyle \frac{\sigma_1}{\mu} = 0.04$}
 \vspace{1.1cm} 
 & \hspace{-0.7cm}
\includegraphics[width=3.8cm]{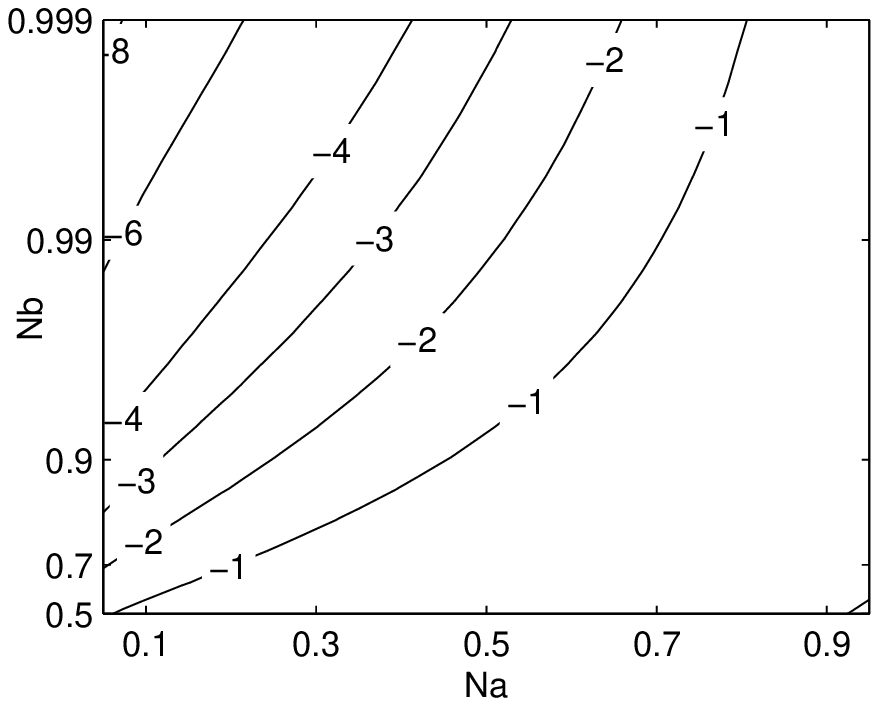}  & \hspace{-0.75cm}
\includegraphics[width=3.8cm]{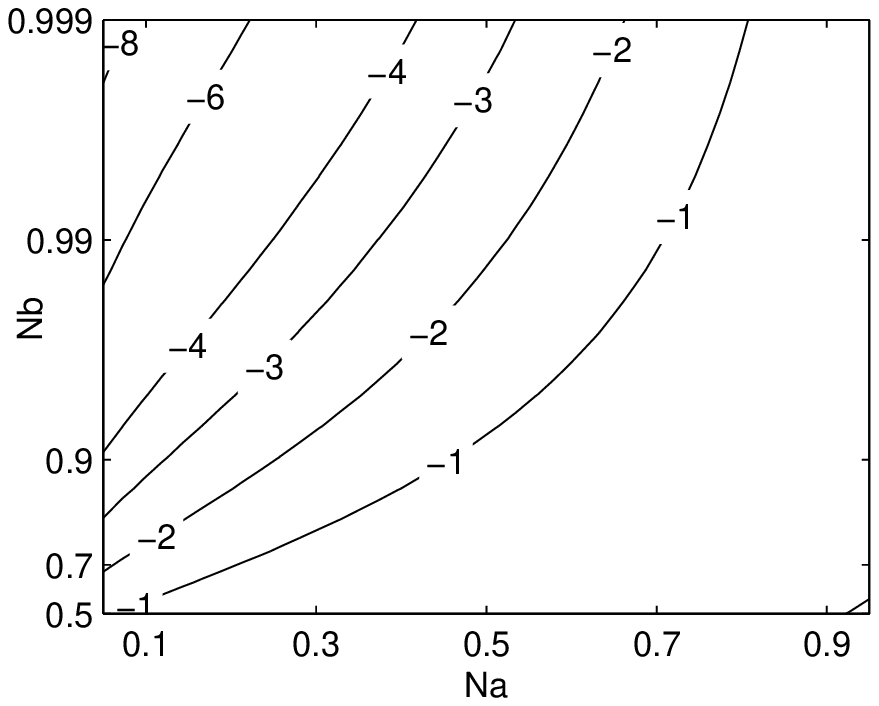}  & \hspace{-0.75cm}
\includegraphics[width=3.8cm]{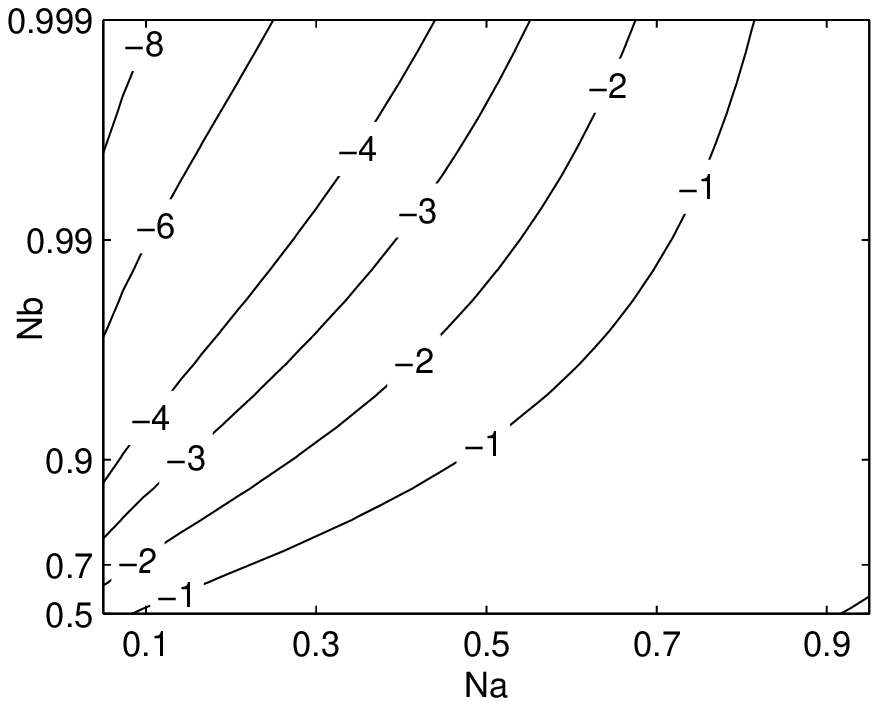}  & \hspace{-0.75cm}
\includegraphics[width=3.8cm]{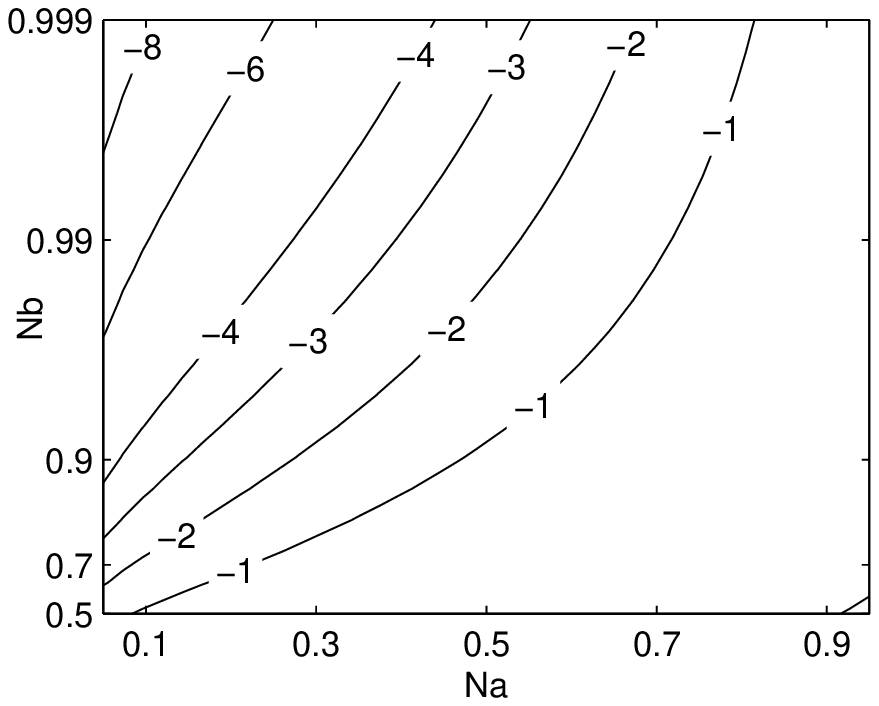}
\end{tabular}
\caption{Expected value of the logarithm of proposal functions affected by the fact that we are using an asymmetrical proposal (see Eq.~(\ref{Eq.terms_AP})) -- first a big jump in the parameter space and then a series of small ones to explore the new region -- to efficiently move between different modes of the target distribution. These results were obtained numerically and they depend on $4$ independent parameters $\{ \sigma_1 / \mu , \sigma_2 / \mu , N_a , N_b \}$ (the panels without any contour lines are because they represent only values greater than $-1$). Since the expected values of the proposals are always smaller than one, we'll refer to them as \emph{the losses} (in the final acceptance probability) \emph{due to an asymmetrical proposal (AP)}.}
\label{Fig.Losses_AP}
\end{figure}

In general, one can compute the expected values of the terms in Eq.~(\ref{Eq.terms_AP}) for the whole parameter space $\{ \sigma_1 , \sigma_2 , \mu , N_a , N_b \}$, trying to quantify the contribution of these ratios of proposal probabilities to the final acceptance probability. In Figure~\ref{Fig.Losses_AP} we plot the results obtained numerically, which are valid in the whole parameter space. An analytical expression can also be derived for the cases where the separation between the Gaussians, $\mu$, is big enough (\ie $\mu >> \sigma_i$):
\beq \label{eq:mean_3gauss}
E \left[ \log \left( \frac{q_a(\beta_i , \beta_{i-1})}{q_a(\lambda , \beta_1)} \; \frac{q_b(\bar{x} [\beta_i , \ldots , \beta_{1} ], \lambda )}{q_b(\bar{x} [\lambda , \ldots , \beta_{i-1}] , \beta_i )} \right) \right]  = (N_a-N_b) \log\left(\frac{\frac{1}{N_a}-1}{\frac{1}{N_b}-1}\right) \; .
\eeq
In this case, the losses in acceptance probability due to an asymmetrical proposal (AP) become independent of $\{ \sigma_1 , \sigma_2 , \mu\}$ as we can see in Figure~\ref{Fig.Losses_AP}, where the $N_a - N_b$ maps tend to a common plot as we go to smaller values of $\sigma_i / \mu$ (low-right plots of the grid). 

\begin{figure}
\centering
\includegraphics[width=6cm]{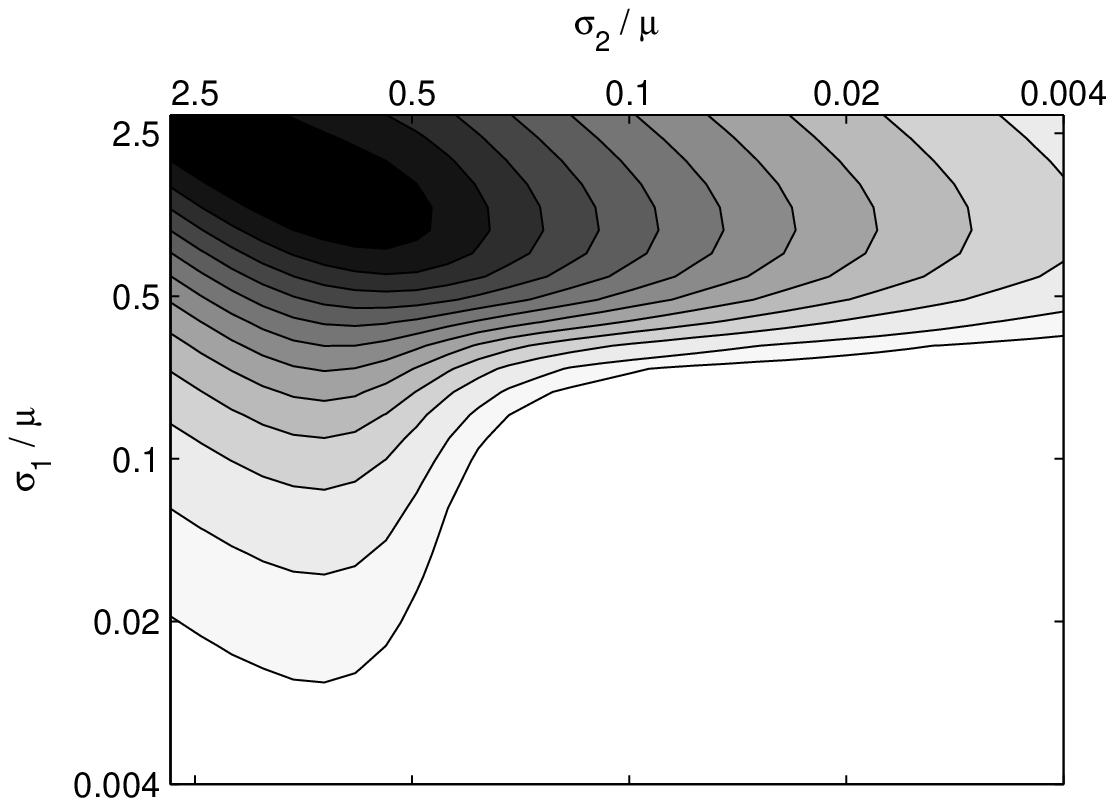}
\caption{Root-mean-squared values of the differences (in a $N_a - N_b$ map, like the ones of Figure~\ref{Fig.Losses_AP}) between the numerical results of Figure~\ref{Fig.Losses_AP} and the analytical expression given by Eq.~(\ref{eq:mean_3gauss}), that was obtained under the assumption that the three Gaussians are well separated. We plot the results as a function of $\sigma_1 / \mu$ and $\sigma_2 / \mu$, getting a quantitative result to evaluate the validity range of Eq.~(\ref{eq:mean_3gauss}).}
\label{Fig.Validity_range_AP}
\end{figure}

The reason why the analytical expression is only valid in the limit of small values of $\sigma_i / \mu$ is due to the assumptions made for its derivation (see Appendix~\ref{ap:Analytical_results} for further details). In Figure~\ref{Fig.Validity_range_AP} we represent the r.m.s. differences of the analytical solution compared with the numerical one in a  $\sigma_1 / \mu - \sigma_2 / \mu$ map, where we can notice that the analytical expression will be applicable when $\sigma_1 / \mu \sim \sigma_2 / \mu \lesssim 0.1$, \ie when the separation between Gaussians is $\gtrsim 10$ times bigger than their typical width.

In Figure~\ref{Fig.Losses_AP}, as well as in Eq.~(\ref{eq:mean_3gauss}), we realise that choosing the solution that the intuition may suggest, with $q_a$ always drawing big jumps (\ie $N_a = 0$) and $q_b$ small ones ($N_b = 1$) would yield an infinite loss\footnote{In the case of Gaussians with a large separation this is strictly true; if the separation is not that big, then the losses are merely huge.} in the acceptance probability. However, from these plots one can see that we can still have a very efficient algorithm by choosing small (big) values of $N_a$ ($N_b$) and getting proposal ratios of the order of $e^{-1} - e^{-2}$, which is acceptable.

\subsection{Choice of proposals: Central location}
\label{subsec:qs_ratio}

The other non-trivial aspect that one should take into account in order to avoid ending up with negligible acceptance probabilities in the DR algorithm is the way that the central location of the mixture of Gaussians proposal, $\bar{x} [\lambda , \ldots , \beta_{i}]$, is chosen based on the preceding sample path. Any of the generic ratios of proposal probabilities that appear in Eq.~(\ref{eq:mastereq}),
\beq \label{eq:terms_to_CPE}
\frac{q_2(\beta_i , \beta_{i-1} , \beta_{i-2})~\ldots ~q_{i-1}(\beta_{i} , \ldots , \beta_1)}
        {q_2(\lambda , \beta_1 , \beta_2)~\ldots ~q_{i-1}(\lambda , \ldots , \beta_{i-1})} 
\quad \longrightarrow \quad
\frac{q_b(\bar{x} [\beta_i , \beta_{i-1}] , \beta_{i-2})~\ldots ~q_b(\bar{x} [\beta_i , \ldots , \beta_2] , \beta_1)}
        {q_b(\bar{x} [\lambda , \beta_1] , \beta_2)~\ldots ~q_b(\bar{x} [\lambda , \ldots , \beta_{i-2}] , \beta_{i-1})}        
\; ,
\eeq
contains the proposal function evaluated over the \emph{reverse chain} in the numerator and evaluated over the \emph{normal chain} in the denominator. The latter term has no issues, because the chain is evaluated with the same probability density function from which it was drawn, but this is not the case for the term in the numerator, where the $\beta_j$ ($1<j<i-1$) element of the chain was generated from a 3-Gaussian function centred at $\bar{x} [\lambda , \beta_1, \ldots, \beta_{j-1}] \equiv \bar{x}_f$ but is evaluated by a function centred at $\bar{x} [\beta_i, \ldots, \beta_{j+1}] \equiv \bar{x}_r$. 

Any difference between these two quantities, $\Delta \equiv \bar{x}_f - \bar{x}_r$, means having a numerator (on average) smaller than the denominator. In particular, in the limit of big separation between Gaussians ($\sigma_i / \mu$ small), one can derive an analytical expression (see Appendix~\ref{ap:Analytical_results}) for the expected value of a single proposal ratio due to a shift, $\Delta$, between the value where the proposal function was centred and the central location of the evaluation function:
\beq \label{eq:mean_centralproposal}
E\left[ \log \left( \frac{q_b(\bar{x} [\beta_i, \ldots, \beta_{j+1}] , \beta_{j})}{q_b(\bar{x} [\lambda , \ldots, \beta_{j-1}] , \beta_j)} \right) \right] = -\frac{\Delta^2}{2} \left( \frac{N_b}{\sigma_1^2} + \frac{1-N_b}{\sigma_2^2} \right) \; .
\eeq

We see that the contribution of each pair of proposals, will always be smaller than $1$ for any $\Delta$, so without a proper way of computing $\bar{x}$, this effect will appear in all the terms inducing a systematic shift of the acceptance probability to very small (negligible) values. Therefore, it is crucial to find an algorithm in which the central location of the proposal is independent from the particular sample we are using to compute $\bar{x}$, or in other words $\bar{x} [\lambda , \beta_1, \ldots, \beta_{j-1}] \simeq \bar{x} [\beta_i, \ldots, \beta_{j+1}], \; \forall j \in (1,i-1)$.

This fact automatically rules out some logical options that one might have in mind as an efficient way to explore the new region of the parameter space, such as (i) using, with a certain probability, the past element of the DR chain with the highest likelihood to draw the next elements, trying to emulate a Metropolis-Hastings algorithm, or (ii) always proposing from the first point in parameter space in which one lands after the `big jump'. On the other hand, we have the opposite case that would be using the last element of the chain to draw the next one, but this would represent a complete blind random walk, and therefore a very inefficient way to explore the parameter space given some prior knowledge of the target distribution. Thus, the exploration of the new region, depicted in Fig.~\ref{Fig.TwoMountains}, has to be performed blindly rather than ``up-hill'', as would be in a standard Metropolis-Hastings algorithm. The real benefits of the DR come from increasing the efficiency of the MCMC algorithm in the \cite{Peskun:1973} sense.

There are several ways to build an algorithm that at the same time explores efficiently the parameter space and yields $\Delta \simeq 0$. The solution we propose here consists in using the mean of all the past elements of the chain after the big jump\footnote{We use all the elements of the DR chain except the first one.} as a central location for the next proposal,
\beq \label{eq:def_central_next_proposal}
\bar{x} [\lambda , \beta_1 , \ldots , \beta_{j-1}] \equiv \mean (\beta_1 , \ldots , \beta_{j-1}) \; ,
\eeq
although, of course, other options may be considered. The main motivation for our decision is that we are using a statistical property of the chain, which has the same expected value independently of the particular sample of elements we are considering\footnote{Of course, given a sample of $N$ elements of a certain random variable, their mean will be closer to the expected value as $N$ increases, so we expect to have bigger $\Delta$'s with the first elements of the chain.}. Moreover, by using the mean value, the generated central proposal values will stabilise near the first point where one lands after the big jump, which is also convenient in order to explore this new region.

\begin{figure}
\hspace{-1.35cm}
\begin{tabular}[t]{p{1.5cm}cccc}
 & \hspace{-0.7cm}
 \scriptsize{ $\displaystyle \frac{\sigma_2}{\mu} = 2.00$} & \hspace{-0.75cm}
 \scriptsize{ $\displaystyle \frac{\sigma_2}{\mu} = 0.54$} & \hspace{-0.75cm}
 \scriptsize{ $\displaystyle \frac{\sigma_2}{\mu} = 0.15$} & \hspace{-0.75cm}
 \scriptsize{ $\displaystyle \frac{\sigma_2}{\mu} = 0.04$}
 \\[0.3cm]
 \vspace{-1.9cm}
 \scriptsize{ $\displaystyle \frac{\sigma_1}{\mu} = 2.00$}
 \vspace{1.1cm}
 & \hspace{-0.7cm}
\includegraphics[width=3.8cm]{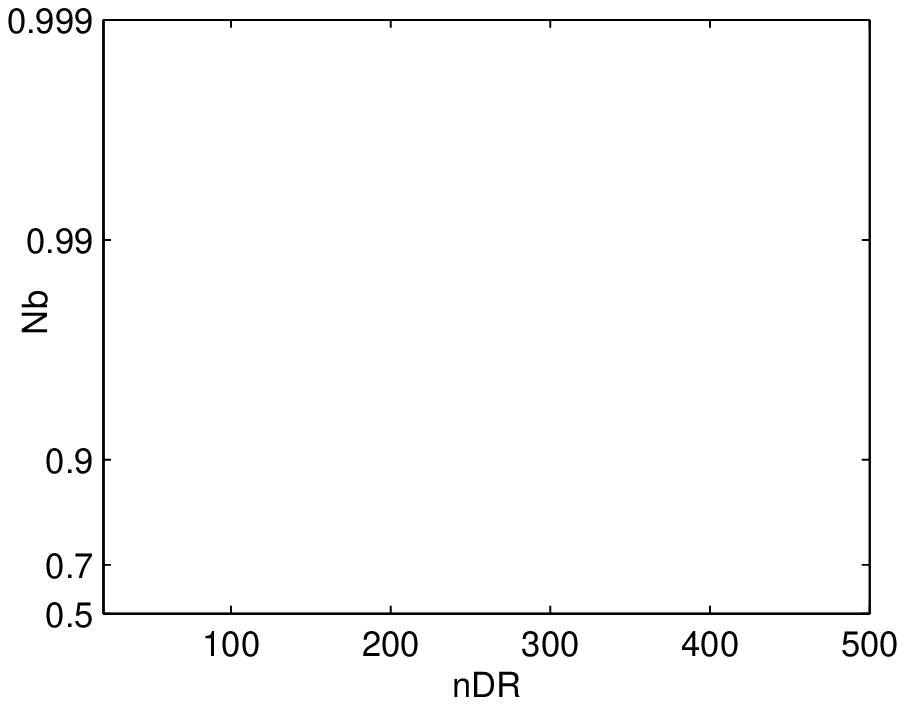} & \hspace{-0.75cm}
\includegraphics[width=3.8cm]{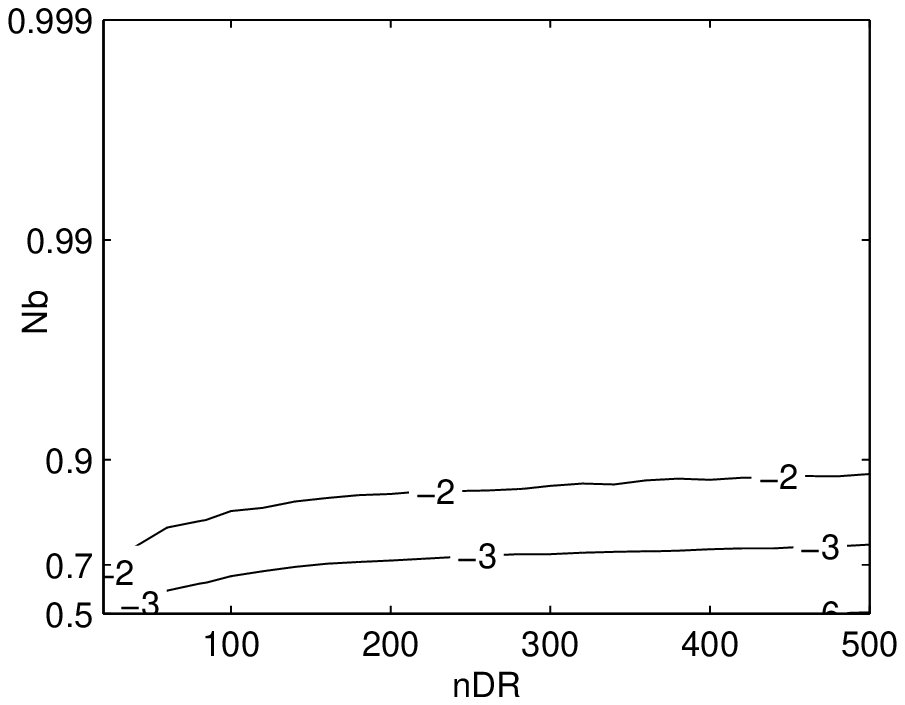} & \hspace{-0.75cm}
\includegraphics[width=3.8cm]{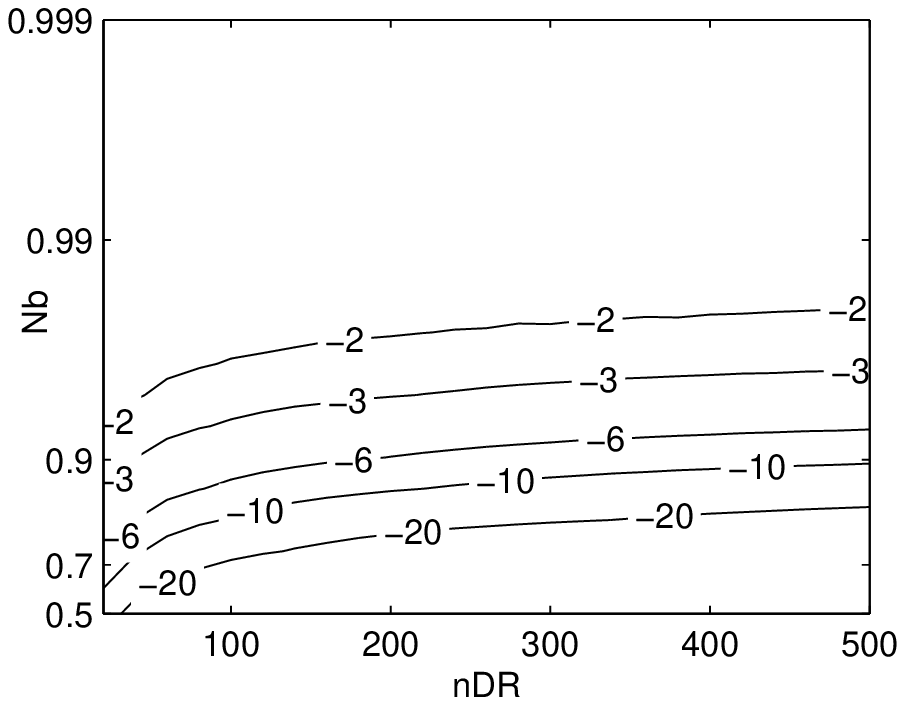} & \hspace{-0.75cm}
\includegraphics[width=3.8cm]{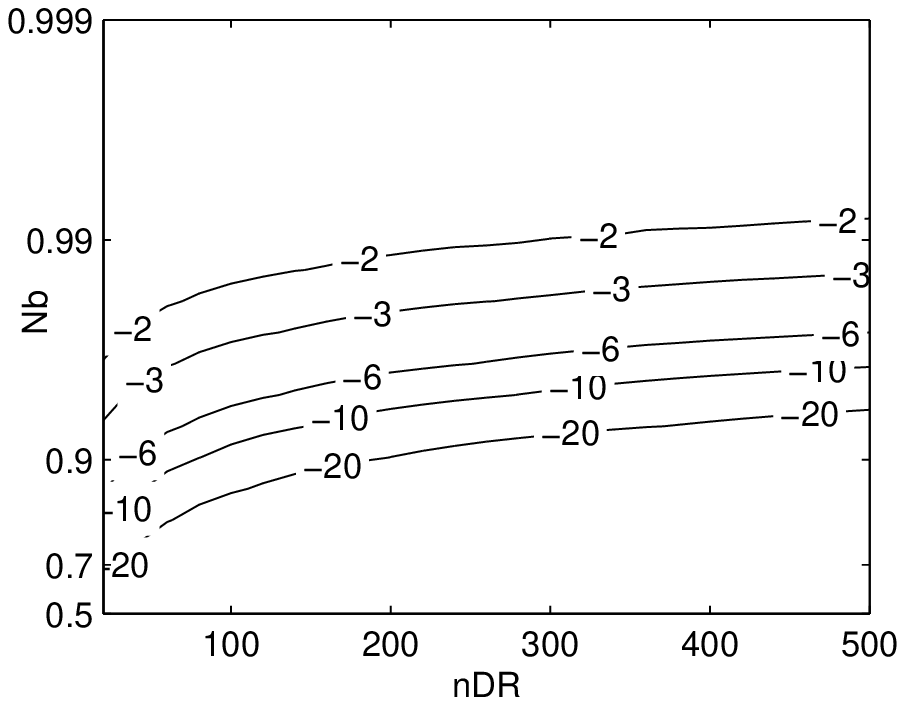}
\\
 \vspace{-1.9cm}
 \scriptsize{ $\displaystyle \frac{\sigma_1}{\mu} = 0.54$}
 \vspace{1.1cm} 
 & \hspace{-0.7cm}
\includegraphics[width=3.8cm]{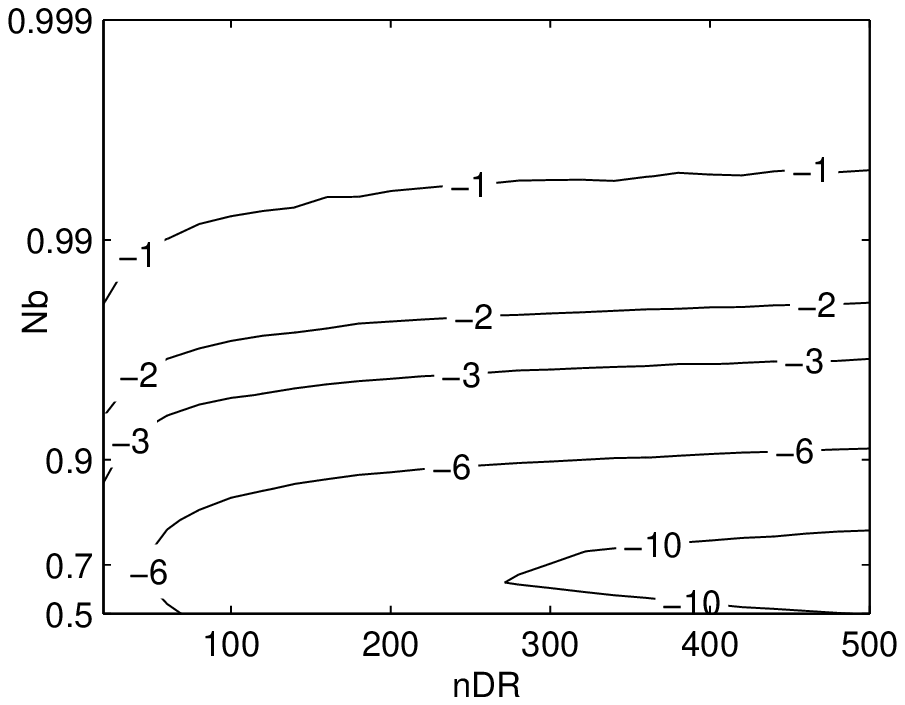} & \hspace{-0.75cm}
\includegraphics[width=3.8cm]{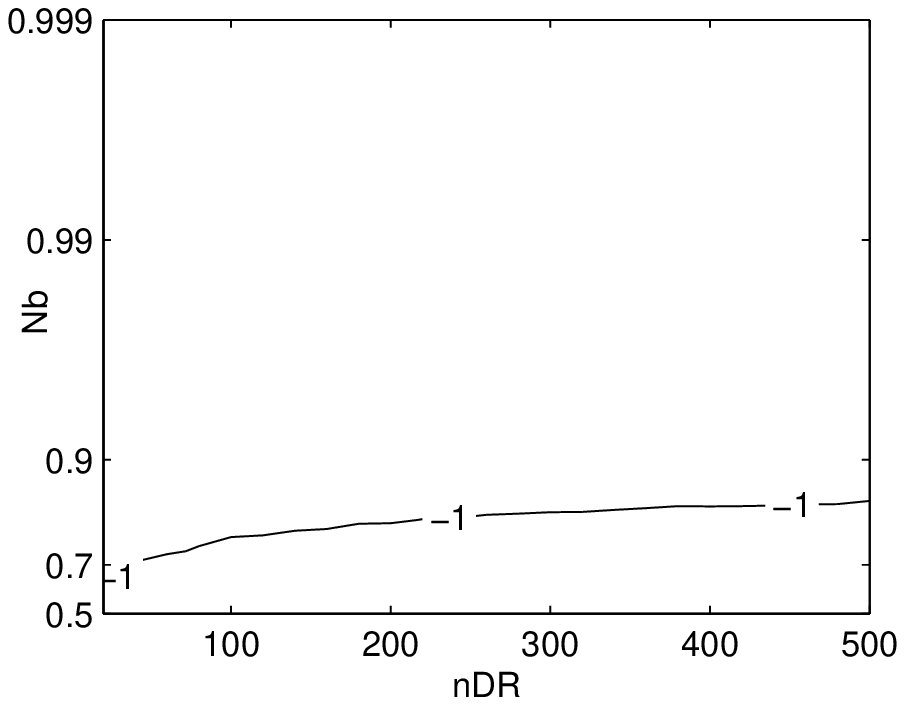} & \hspace{-0.75cm}
\includegraphics[width=3.8cm]{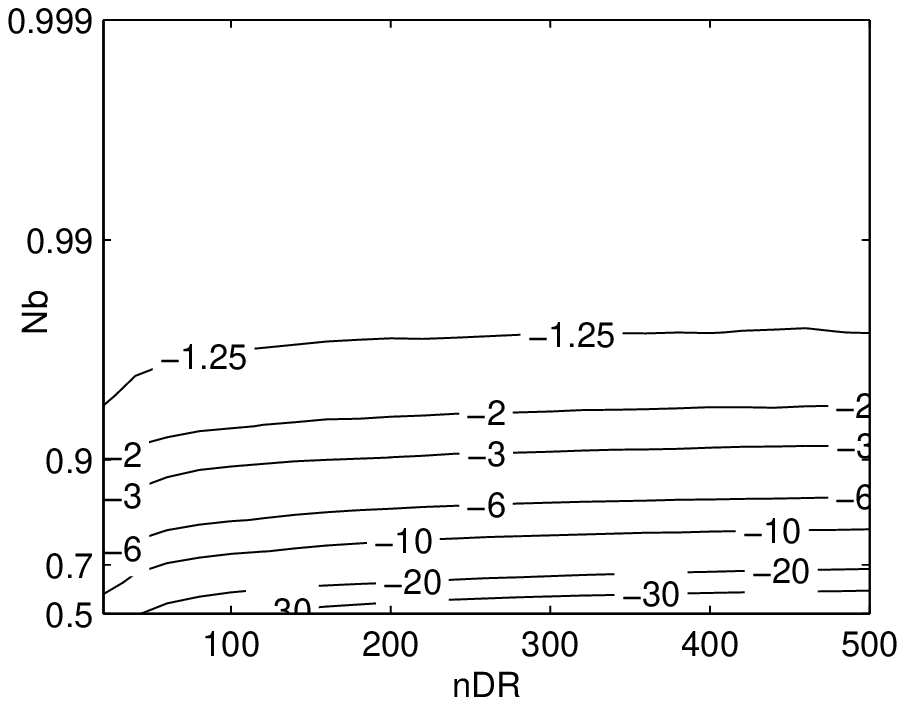} & \hspace{-0.75cm}
\includegraphics[width=3.8cm]{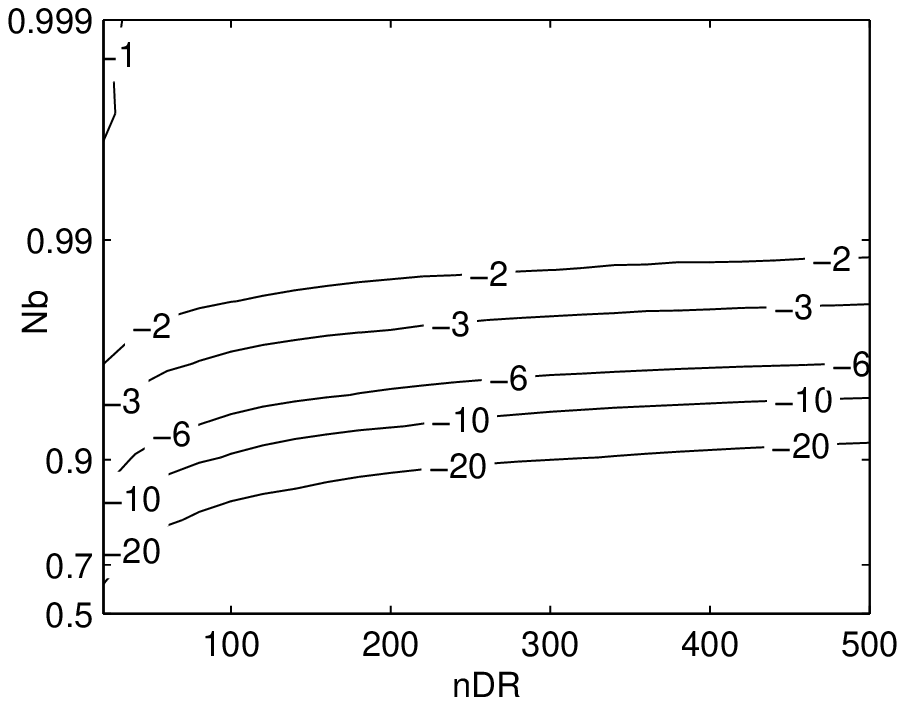}
\\
 \vspace{-1.9cm}
 \scriptsize{$\displaystyle \frac{\sigma_1}{\mu} = 0.15$}
 \vspace{1.1cm} 
 & \hspace{-0.7cm}
\includegraphics[width=3.8cm]{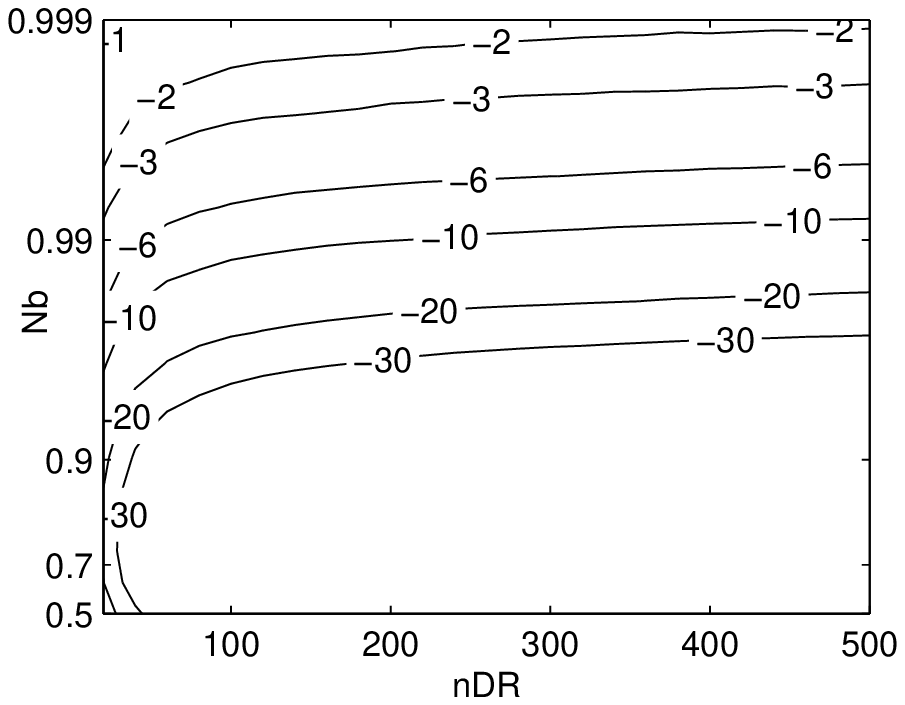}  & \hspace{-0.75cm}
\includegraphics[width=3.8cm]{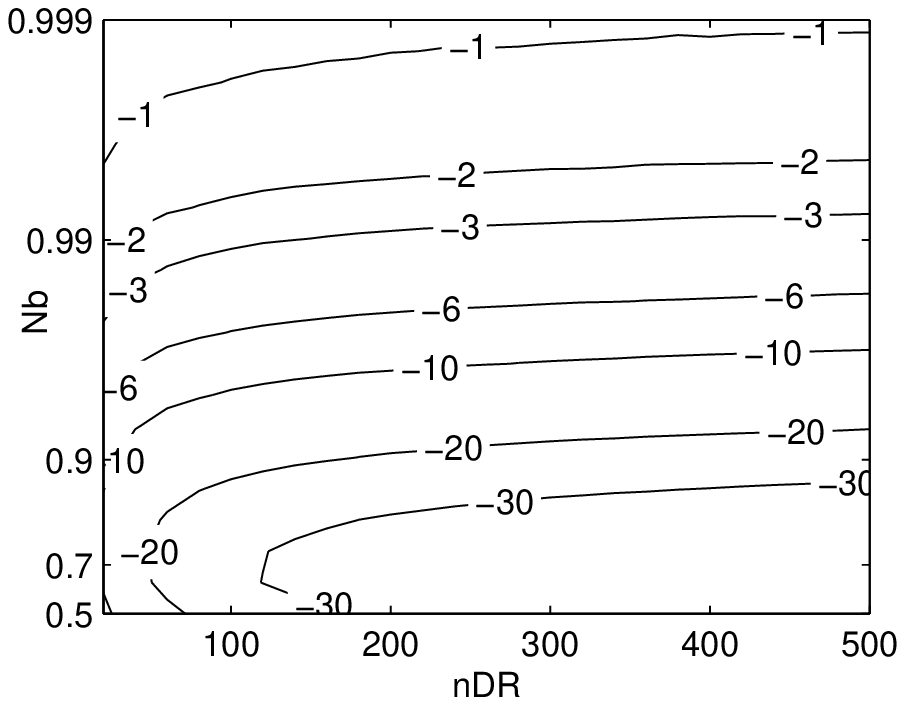}  & \hspace{-0.75cm}
\includegraphics[width=3.8cm]{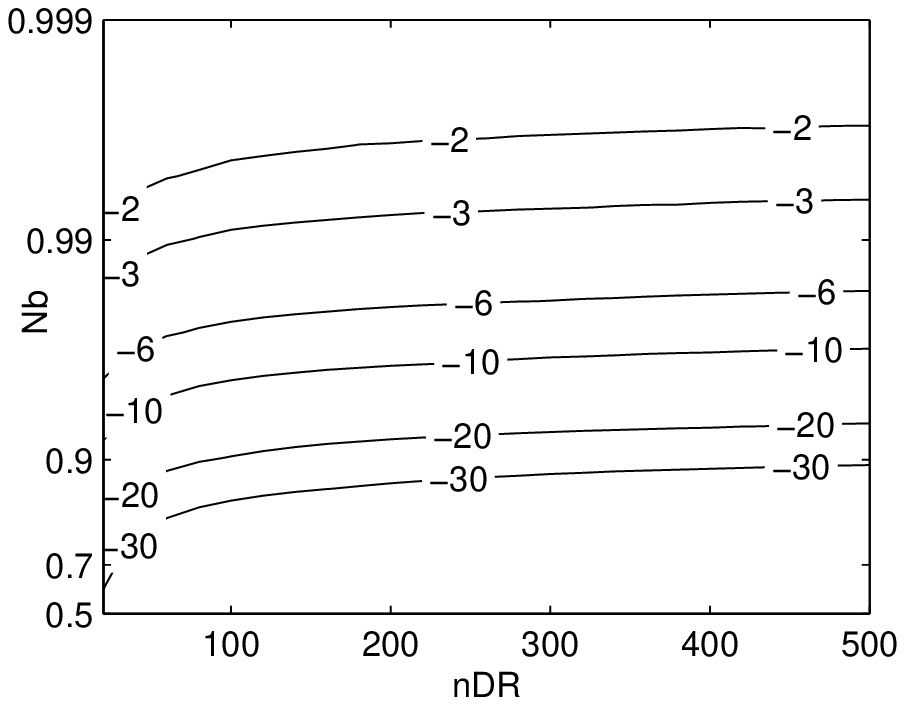}  & \hspace{-0.75cm}
\includegraphics[width=3.8cm]{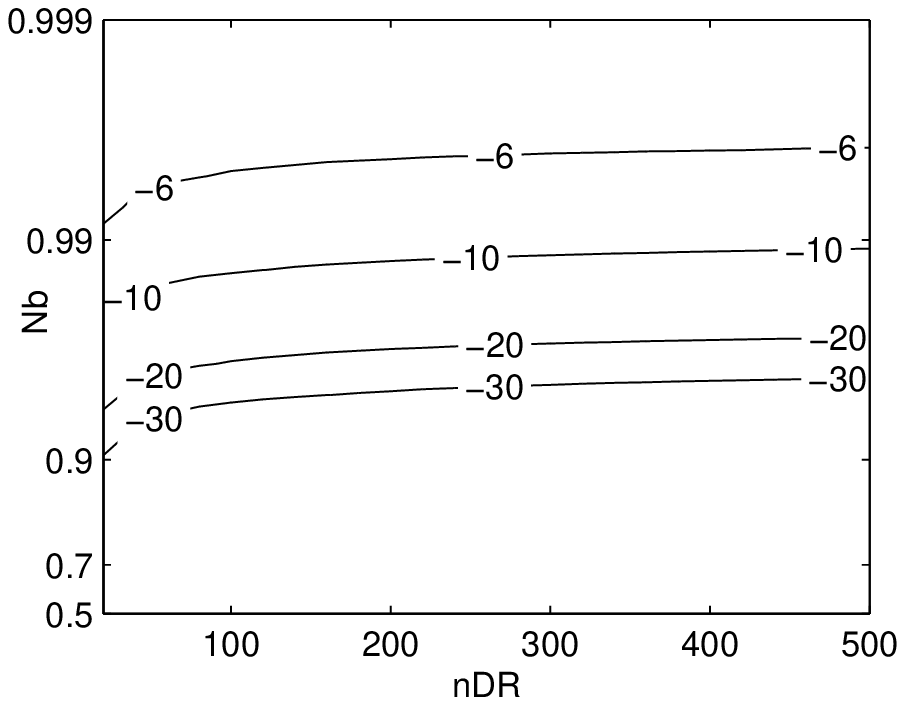}
\\
 \vspace{-1.9cm}
 \scriptsize{$\displaystyle \frac{\sigma_1}{\mu} = 0.04$}
 \vspace{1.1cm} 
 & \hspace{-0.7cm}
\includegraphics[width=3.8cm]{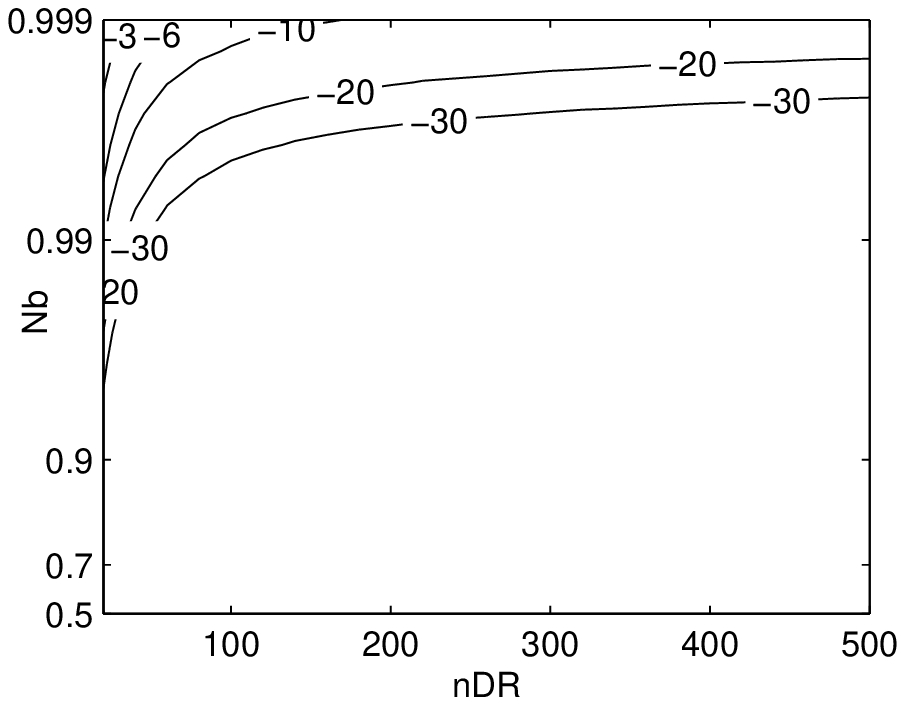}  & \hspace{-0.75cm}
\includegraphics[width=3.8cm]{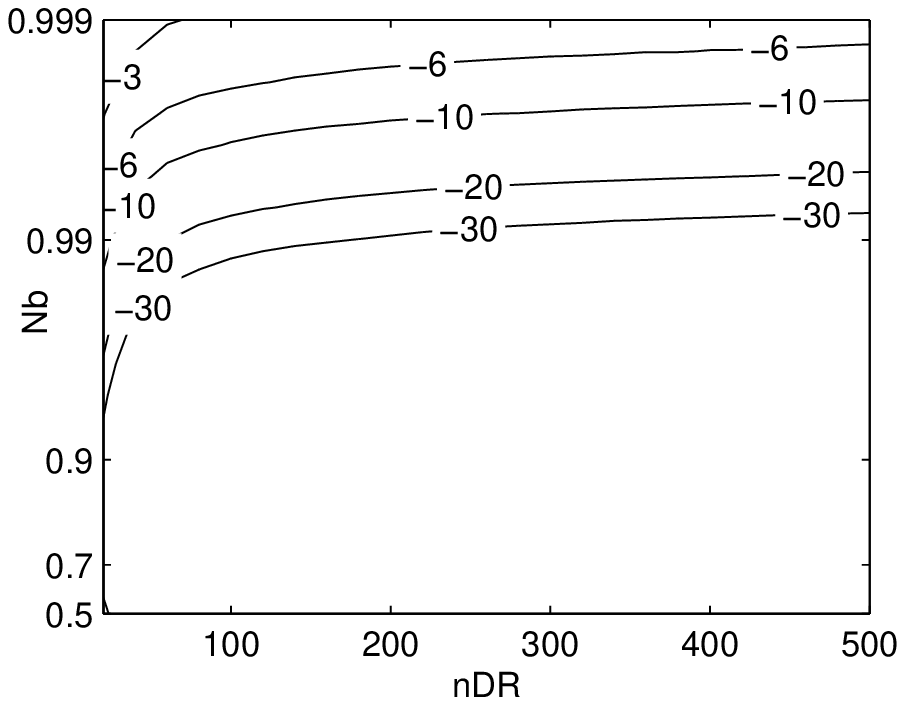}  & \hspace{-0.75cm}
\includegraphics[width=3.8cm]{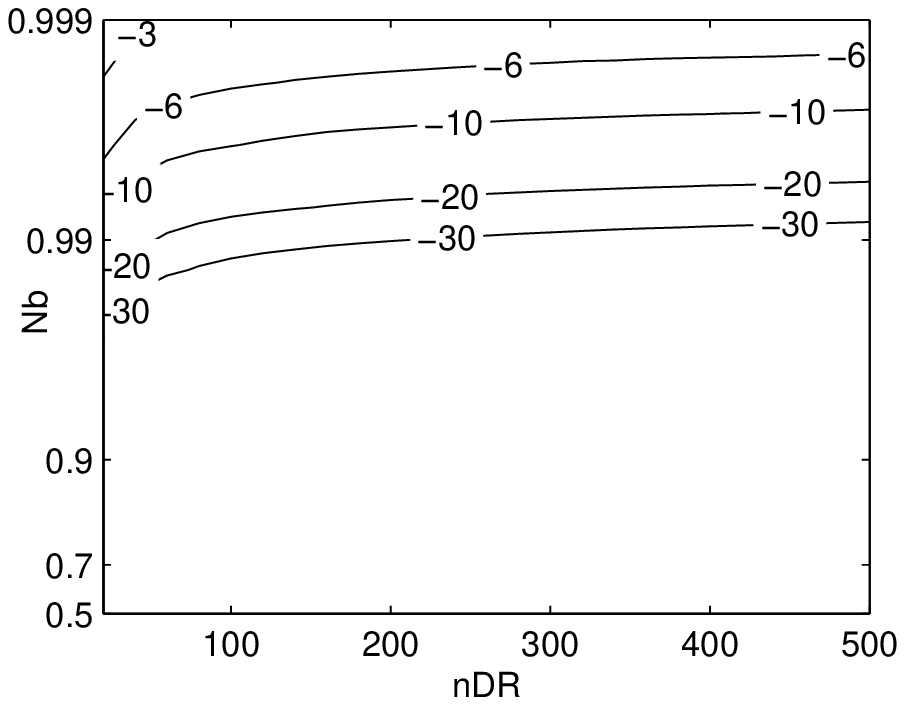}  & \hspace{-0.75cm}
\includegraphics[width=3.8cm]{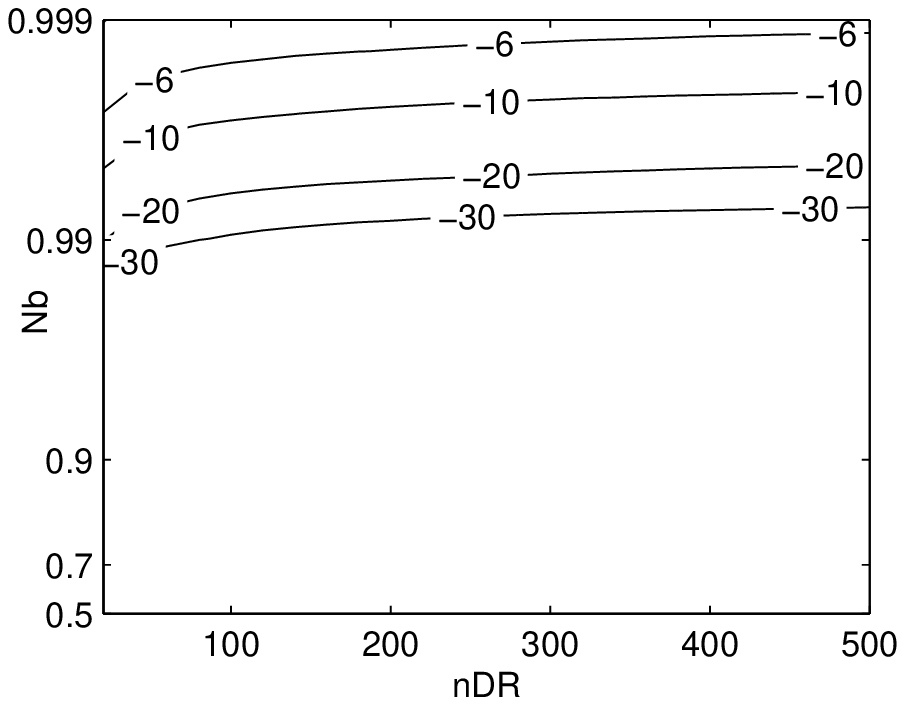}
\end{tabular}
\caption{Expected value of the logarithm of proposal functions from Eq.~(\ref{eq:terms_to_CPE}) obtained numerically for the whole parameter space, $\{ \sigma_1 / \mu , \sigma_2 / \mu , N_b , \nDR \}$, using the same kind of representation as in Figure~\ref{Fig.Losses_AP}. These results are obtained after having made the decision of computing $\bar{x} [\lambda , \ldots , \beta_i ]$ as the mean value of all the elements of chain after the big jump (see Eq.~(\ref{eq:def_central_next_proposal})). The reason why the expected values are always smaller than $1$ is because the central location of the proposals evolves; that's why we will refer to these results as \emph{the losses due to the central proposal evolution (CPE)}.}
\label{Fig.Losses_CPE}
\end{figure}

Once we have made this choice, one can numerically compute the expected values of the proposals ratio of Equation~(\ref{eq:terms_to_CPE}) (\ie losses in the final acceptance probability due to the central proposal evolution (CPE)) in the whole parameter space, here defined by $\{ \sigma_1 , \sigma_2 , \mu , N_b , \nDR \}$, where $\nDR$ is the number of elements of the DR chain. In Figure~\ref{Fig.Losses_CPE}, we plot these results in the same way as in Sec.~\ref{subsec:3_gauss}, \ie a grid in the $\sigma_1 / \mu - \sigma_2 / \mu$ space of maps in the $N_b - \nDR$ plane. On this occasion it was not possible to get an analytical expression as we justify in Appendix~\ref{ap:Analytical_results}.

By looking at Figure~\ref{Fig.Losses_CPE} one can see that the losses due to CPE are very sensitive to the actual values of $\sigma_i / \mu$. When the Gaussians have some separation (\ie $\sigma_i / \mu$ becomes small), we need very large values of $N_b$ in order not to obtain negligible contributions to the acceptance probability. The dependency on the number of stages that we consider in a Delayed Rejection, $\nDR$, is not that important: at the beginning, $\nDR \lesssim 100$, the losses increase as we add more steps to the DR chain, but after that first period the values stabilise ending up almost constant; in fact, Figure~\ref{Fig.Losses_CPE} is plotted only up to $\nDR = 500$, although we checked that the contour lines are almost constant for $\nDR > 500$.

\subsection{Delayed Rejection in practice: Choice of parameters}
\label{subsec:parameters}

In the previous sections we have discussed all the possible problems that one can find trying to use the Delayed Rejection algorithm as an efficient method to explore the local maxima of the target distribution, and their solutions. The purpose of this subsection is to give the general scheme of how to proceed in a real problem and how to make the selection of parameters in light of our results, in order to get an efficient algorithm with non-negligible acceptance probabilities.

We use the general DR algorithm introduced by \cite{TierneyMira:1999, Mira:2001}, choosing as proposal probability functions two kinds of 3-Gaussian distributions (one $q_a$ to propose an initial big jump in the parameter space trying to land in the neighbourhood of a new maximum, and an arbitrary number of $q_b$ to explore this new region where we landed) parametrised according to Eq.~(\ref{eq:Proposals}) and graphically represented in Figure~\ref{Fig.Proposals}. This gives a set of six parameters, $\{ \sigma_1 , \sigma_2 , \mu , N_a , N_b , \nDR \}$ that we are free to tackle at best the problem at hand. The way to proceed should be the following:
\begin{enumerate}
\item The first three parameters, $\{ \sigma_1 , \sigma_2 , \mu \}$, are chosen to have a proposal adapted to the multimodal structure of the target distribution, $\mu$ being the typical distance between two maxima of the target distribution and $\sigma_i$ their typical width. So a rough previous knowledge of the target is required.
\item Once we know the first three parameters, one can go to the plots of Figures~\ref{Fig.Losses_AP} and \ref{Fig.Losses_CPE} and look at the maps corresponding to the closest $\sigma_1 / \mu - \sigma_2 / \mu$ values they have. In particular, one first can look at \emph{losses due to the central proposal evolution (CPE)} (Figure~\ref{Fig.Losses_CPE}) to select an $N_b$ value high enough\footnote{The tolerance range depends on the typical ratio between the likelihood of the main maximum and the likelihood of the secondary ones. In our experience, allowing losses of the order of $e^{-3} - e^{-4}$ is acceptable.} to have acceptable expected values of this ratio of proposals. Normally one works with $\nDR \gtrsim 100$, where the results are almost independent of this parameter. 
\item After knowing the appropriate $N_b$ value to use, one has to look at the \emph{losses due to an asymmetrical proposal (AP)} (Figure~\ref{Fig.Losses_AP}) in order to get the value of $N_a$. Although in principle, $N_a$ should be as small as possible in order to mostly make big jumps at the first step, sometimes\footnote{Mostly when we are facing a problem with highly separated maxima, that requires high values of $N_b$ in order to avoid huge losses due to CPE.} it is necessary to choose larger values in order to reduce the losses in AP. Having a large $N_a$ value means that not all the DR chains will start with a big jump in the parameter space; this is however not a problem because although sometimes the DR chain will initially explore the same maximum in which the chain is, this is desirable once the global maximum is reached.
\item Since the dependency with $\nDR$ is very weak, the criterion to choose this last parameter will be mainly driven by the constraints on the computational costs. 
\end{enumerate}


\section{Numerical Implementation}
\label{sec:implementation_issues}

The Delayed Rejection algorithm has been applied to many problems \citep{Al-Awadhi:2004, Umstatter:2004, Raggi:2005, Harkness:2000, Robert, Haario:2006}, but this is the first time to our knowledge that it has been implemented in its general scheme with an arbitrary number of stages. In this section we focus on practical numerical implementation issues that one faces in applying a general DR scheme (with no simplifications) with $\nDR$ steps.


\subsection{Efficient computation of the transition probabilities}

\label{subsec:computing_all_terms}

The general expression for the acceptance probability of the \capN -th stage ($\capN \le \nDR$) of the DR, see Eq.~(\ref{eq:mastereq}), can be rewritten following \cite {Mira:2001} as $\alpha_\capN (\lambdav , \betav_1 , \ldots , \betav_\capN ) \equiv 1 \wedge \frac{N_\capN}{D_\capN}$. Since the denominator contains the terms evaluating the chain in the natural order, its value after adding a new stage can be computed recursively, reusing many of information from the previous stage, 
\beq \label{eq:DN}
D_\capN = D_{\capN -1} ~ q_\capN (\lambdav , \betav_1 , \ldots , \betav_\capN ) \, \left[ 1 - \alpha_{\capN -1}( \lambdav, \ldots \betav_{\capN -1}) \right] \; .
\eeq

On the other hand, all the terms of the numerator must be computed at each step,
\beq \label{eq:NN}
N_\capN = \pi( \betav_\capN ) ~ q_1(\betav_\capN , \betav_{\capN -1}) \ldots q_\capN(\betav_\capN , \ldots , \lambdav) ~ \left[ 1 - \alpha_1(\betav_\capN , \betav_{\capN -1} ) \right] \ldots \left[ 1 - \alpha_{\capN -1}(\betav_\capN , \ldots , \betav_1) \right] \; .
\eeq
Furthermore, since the \capN-th acceptance probability involves $\capN-1$ $\alpha$'s, unless we reuse the results of previous iterations of the DR chain, we would have to compute $2^{\capN-1}$ additional acceptance probabilities for the numerator of the \capN-th stage.

Fortunately, most of the terms mentioned above are repeated or computed in previous stages of the DR chain, and only $\capN$ new terms have to be computed at the \capN-th stage. Here we describe a way to efficiently implement the algorithm so as to avoid unnecessary computations, resulting in the maximum efficiency. Our procedure is based on a property of the \capN -th term of the acceptance probability that relates the result from evaluating the chain in the forward and reverse orders,
\beq \label{eq:prop_alphas}
\frac{\alpha_\capN (\lambdav, \ldots , \betav_\capN )}{\alpha_\capN (\betav_\capN , \ldots , \lambdav)} = \frac{N_\capN }{D_\capN } \; .
\eeq

\begin{figure}
\begin{center}
\includegraphics[width=\textwidth]{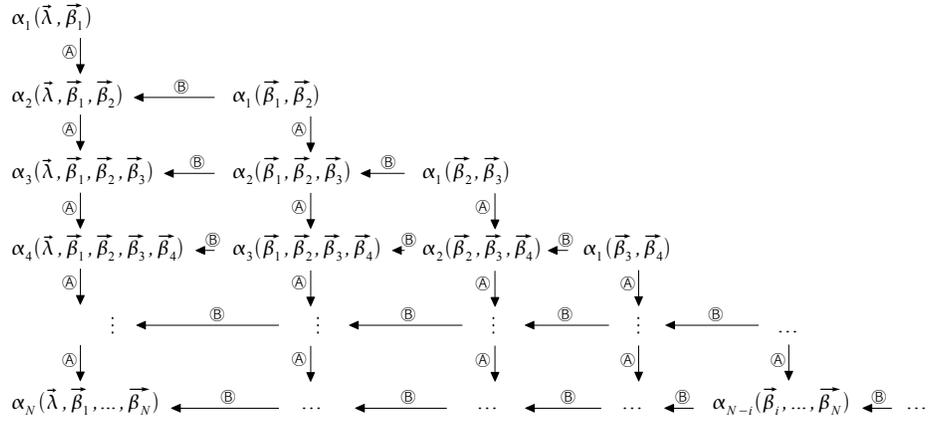}
\end{center}
\caption{Example of the table that need to be built in order to efficiently implement a general Delayed Rejection algorithm with an arbitrary number of stages, $\nDR$. Only the terms in the first column represent the actual acceptance probabilities of the DR chain, but all the others are required in order to compute those first ones. The table must be constructed by rows, starting each row with the element at the rightmost  end. For an arbitrary element of the table, its denominator can be computed reusing the calculations of the element just above (see Eq.~(\ref{eq:DN})) (A), and for the numerator we will use the previous calculations done for all the elements at its right hand side (see Eq.~(\ref{eq:NN}) together with Eq.~(\ref{eq:prop_alphas})) (A). This is the most efficient way to proceed in the implementation of a general DR algorithm, reducing the number of new quantities to be computed at the new \capN -th stage from $2^{\capN -1}$ (in the most inefficient scenario) to $\capN$.}
\label{Fig.table_alphas}
\end{figure}

From the results of Eq.~(\ref{eq:DN}) and (\ref{eq:prop_alphas}), one can build a table like the one shown in Figure~\ref{Fig.table_alphas} a row at a time, starting each row from the right hand side.
By making use of Equation~(\ref{eq:DN}), the denominator of any element of the table can be computed using the information from the element just one position up (A); and the numerator will involve all the elements at its right hand side with the parameters in the reverse order (B); here is where we make use of Eq.~(\ref{eq:prop_alphas}). Of course, only the elements of the leftmost column will correspond to the real acceptance probabilities of a certain stage.

One extra benefit from always working with chains in the `normal' order, and just using Eq.~(\ref{eq:prop_alphas}) to compute the chains with `reverse' order is that we can easily reuse proposal probability values computed at previous elements. This already was the case for the denominator as we can see in Eq.~(\ref{eq:DN}), which contains only a single new proposal probability to compute. For the numerators it turns out that the proposal probabilities are repeated as we move left in the table of Figure~\ref{Fig.table_alphas}, so only a single value needs to be computed for each new element of the table.

This procedure is completely general and independent of the proposal distributions that are used. In terms of computational costs, we can see from Figure~\ref{Fig.table_alphas} that building a DR chain with $\capN$ stages will involve the computation of $\approx \frac{1}{2} (\capN ^2+\capN)$ terms, whereas a normal chain would have involved $\sim \capN$. Of course, in the cases that one can simplify the problem working with proposal functions that do not depend on all the past elements of the chain, then the general scheme can be simplified further, giving a faster algorithm as can be seen in all the previous applications of the Delayed Rejection method. 

\subsection{Avoiding numerical divergences}
\label{subsec:alphas}

The general expression for the acceptance probability at the \capN -th stage of the Delayed Rejection, Eq.~(\ref{eq:mastereq}), contains $2(\capN -1)$ factors that correspond to $(1-\alpha)$'s. Since $\alpha_\capN = 1 \wedge \frac{N_\capN}{D_\capN}$, it is possible that some of these terms could be zero\footnote{In this case, the asymptotic behavior of the numerator and denominator is the same, so their ratio would be finite.}.
In a real implementation, this fact must be taken into account when calculating the acceptance probability in the Delayed Rejection chain. In this section we will discuss this apparent problem, and show that a finite acceptance probability is always found.

In addition to Eq.~(\ref{eq:prop_alphas}), another conclusion that can be drawn from comparing the acceptance probabilities of the forward and reverse versions of one chain, is that if one of them is not equal to $1$, the other must be $1$. Taking this into account and looking at the $(1-\alpha)$ terms that appear in Eq.~(\ref{eq:mastereq}), one can see that:
\bea \label{eq:num_zeros}
\# ~zeros~in~D_\capN       &       =       &       \# \left[ \alpha_{j - i}(\betav_i, \ldots , \betav_j) = 1 \right] \, , \quad i < j < \capN\,,  \nonumber \\
\# ~zeros~in~N_\capN       &       =       &       \# \left[ \alpha_{\capN - j}(\betav_j , \ldots , \betav_\capN) \neq 1 \right] \, , \quad i < j < \capN\,.
\eea
In other words and in terms of the procedure explained in Figure~\ref{Fig.table_alphas}, this means that the number of zeros in the denominator of a certain acceptance probability will be equal to the number of $\alpha$s above it in the diagram that are equal to one; whereas the number of zeros in the numerator will be equal to the number of $\alpha$s on its right hand side not equal to unity.

Making use of Equations~(\ref{eq:num_zeros}), and focusing on the actual acceptance probabilities in the leftmost column of Figure~\ref{Fig.table_alphas}, we can see that:
\begin{itemize}
\item The number of zeros in denominator will be zero, because if it were not the case, then a previous stage of the Delayed Rejection would have been accepted.
\item Therefore, in order to have a non-zero acceptance probability, all the $\alpha$s to the right at the same row of the diagram of Figure~\ref{Fig.table_alphas} must be one.
\end{itemize}

To show that the second condition is indeed possible, we will take as example the case of the third stage acceptance probability, $\alpha_3(\lambdav, \betav_1, \betav_2, \betav_3)$. We have already established that the $\alpha_1(\lambdav,\betav_1)$ and $\alpha_2(\lambdav,\betav_1,\betav_2)$ above cannot be one, so we shall show that the entries to the right can be one, starting with $\alpha_1(\betav_2,\betav_3)$. 
\begin{itemize}
\item As $\alpha_1(\betav_2,\betav_3)= \frac{\pi(\betav_3)q_1(\betav_3,\betav_2)}{\pi(\betav_2)q_1(\betav_2,\betav_3)}$, and the $q$ functions are symmetrical in the order of their arguments, we only need $\pi(\betav_3) \ge \pi(\betav_2)$ in order to have an acceptance probability equal to one.

\item Next, the term $\alpha_2(\betav_1,\betav_2,\betav_3)$ can be equal to one, in either of cases
\begin{itemize}
\item Any of the terms above it, in this case just $\alpha_1(\betav_1,\betav_2)$, is one, which would make the denominator equal to zero. In this case we take the minimum of the ratio and 1, which is of course 1.
\item Otherwise, the ratio is finite, and can be $\ge{}1$.
\end{itemize}
\end{itemize}

Therefore, we see in this example that both of the terms to the right can be one, and therefore that $\alpha_3(\lambdav,\betav_1,\betav_2,\betav_3)$ can be non-zero. From Figure~\ref{Fig.table_alphas} it is easy to see that the same argument applies at any stage of the DR. So we see that despite the possibility of zeros existing in the intermediate terms in Figure~\ref{Fig.table_alphas}, the final acceptance probability at all stages of the Delayed Rejection chain is finite, as we desire.



\section{An application: Analysis of gravitational-wave data}

\label{sec:application}

In this section we present results obtained by applying the Delayed Rejection algorithm described above to a specific problem: the search for gravitational waves emitted by short-period close compact binary systems in the data of the Laser Interferometer Space Antenna (LISA) \citep{Bender:1998}. LISA is a space based laser interferometer to survey the sky in the gravitational-wave observational window and is expected to observe a variety of sources, especially binary systems: galactic close stellar-mass compact objects; high-redshift massive black hole binaries; and also stellar mass black holes or degenerate stars orbiting a massive black in the centre of galactic nuclei. The mission is currently being developed by ESA and NASA with an expected launch date in the time frame 2018+. In essence, the LISA data set consists of three time series that contain full information about all the sources, as the instrument is an all-sky monitor and the expected sources are, for the overwhelming majority, long-lived compared to the mission lifetime; so the signals are always ``on'' and overlap with each others. Most of the signals will be buried in the noise, and optimal filtering needs to be applied to positively identify them. The analysis of the LISA data provides an interesting challenge to disentangle an unknown number (of the order of $\sim 10^4$) of signals overlapping in time and frequency, and superimposed to the instrumental noise and foreground radiation produced by the brightness of the gravitational wave sky in this frequency region. Every individual signal is described by $\approx 10$ parameters, and therefore one looks for fitting $\sim 10^5$ parameters out of $\approx 10^7$ data points. 

Bayesian methods have proven to be particularly promising in tackling these issues, and a vigorous effort  is currently on-going to develop these techniques for essentially the whole range of signals that one expects that LISA will be able to observe. Due to the specific nature of the signals and the instrument motion during the observation time, the target distribution shows many secondary maxima and makes the problem of efficiently and robustly computing the (marginalised) posterior density function particularly hard. A considerable variety of Monte Carlo integration methods based on the Metropolis-Hastings acceptance criterion have been developed, and most of them rely crucially on \emph{ad-hoc} strategies to probe multimodal target functions, for stellar mass binary systems \citep{Crowder:2007}, massive black hole binaries \citep{Cornish:2007} and extreme-mass ratio inspirals \citep{Cornish:2008, Gair:2008, Babak:2009}; although recently, parallel tempering techniques have shown a good performance sampling the target distribution from a single stellar mass binary system \citep{Littenberg:2009}.  This is an ideal arena in which a Delayed Rejection scheme would provide significant advantages to MCMC analysis algorithms.

In short, the test analysis that we consider here can be summarised as follows. We have a data set
\beq
d(t) = n(t; \vec{\zeta}_\mathrm{n}) + h(t;\vec{\zeta}_\mathrm{s})
\label{e:da}
\eeq
which consists of the superposition of noise $n(t; \vec{\zeta}_\mathrm{n})$ and gravitational-wave signal $h(t;\vec{\zeta}_\mathrm{s})$, described by a vector of unknown parameters $\vec{\zeta}_\mathrm{n}$ and $\vec{\zeta}_\mathrm{s}$, respectively. We will indicate with $\vec{\zeta} = \{\vec{\zeta}_\mathrm{n}, \vec{\zeta}_\mathrm{s}\}$ the vector of all the $M$ unknown parameters that describe the problem. By applying the Bayes' theorem, we can compute the \emph{joint} posterior density function (PDF), $p(\vec{\zeta}| d)$, of $\vec{\zeta}$ given the data sets, $d$,
\beq
p(\vec{\zeta}|d) = \frac{p(\vec{\zeta}) \, \mathcal{L}(d | \vec{\zeta})}{p(d)}\,,
\label{e:pdf}
\eeq
where $\mathcal{L}(d | \vec{\zeta})$ is the likelihood function and $p(\vec{\zeta})$ the prior probability density function; $p(d)$ is the marginal likelihood or evidence that for the problem at hand plays simply the role of a normalisation constant. Given a sub-set of parameters, say $\zeta_1,\dots,\zeta_m$, one wants to compute the marginalised posterior density function:
\beq
p(\zeta_1,\dots,\zeta_m|d) = \int d\zeta_{m+1}\,\dots\int d\zeta_\mathrm{M} \,
\frac{p(\vec{\zeta}) \, \mathcal{L}(d | \vec{\zeta})}{p(d)}\,.
\label{e:pdfmarg}
\eeq
The case that we consider here is the simplest one for the LISA analysis problem: the computation of the posterior density function of the parameters that characterise a single stellar mass compact binary system in Gaussian and stationary data of unknown spectral density. However, this specific problem provides all the conceptual issues that we will be presented in a real analysis and more complex analysis tasks, such as the search and computation of the posterior density function for (spinning) black hole binary systems and extreme-mass ratio inspirals. 

\begin{figure}
\begin{center}
\begin{tabular}{cc}
\hspace{-0.65cm} \includegraphics[height=7cm]{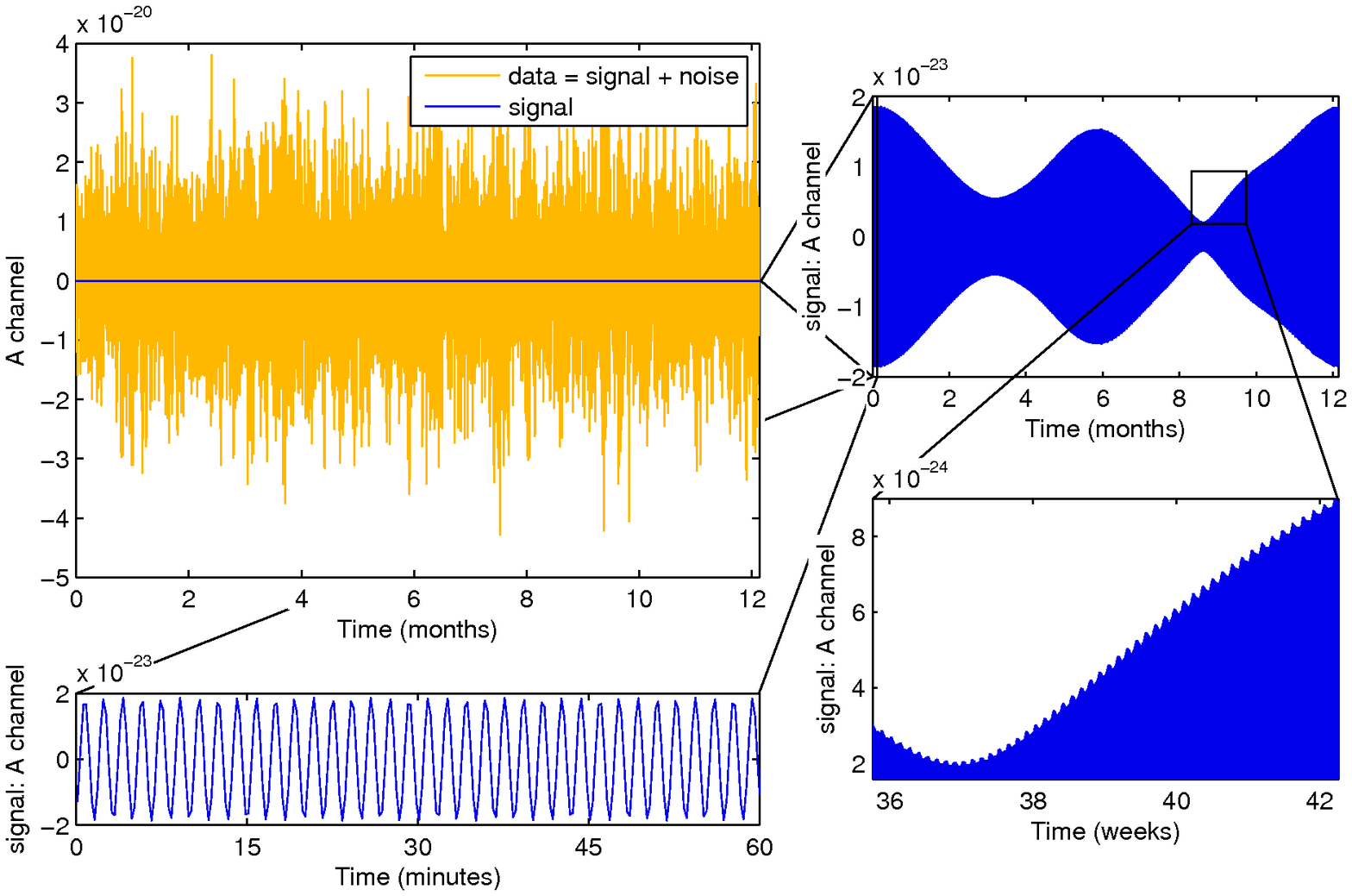} & \hspace{-0.5cm} \includegraphics[height=7cm]{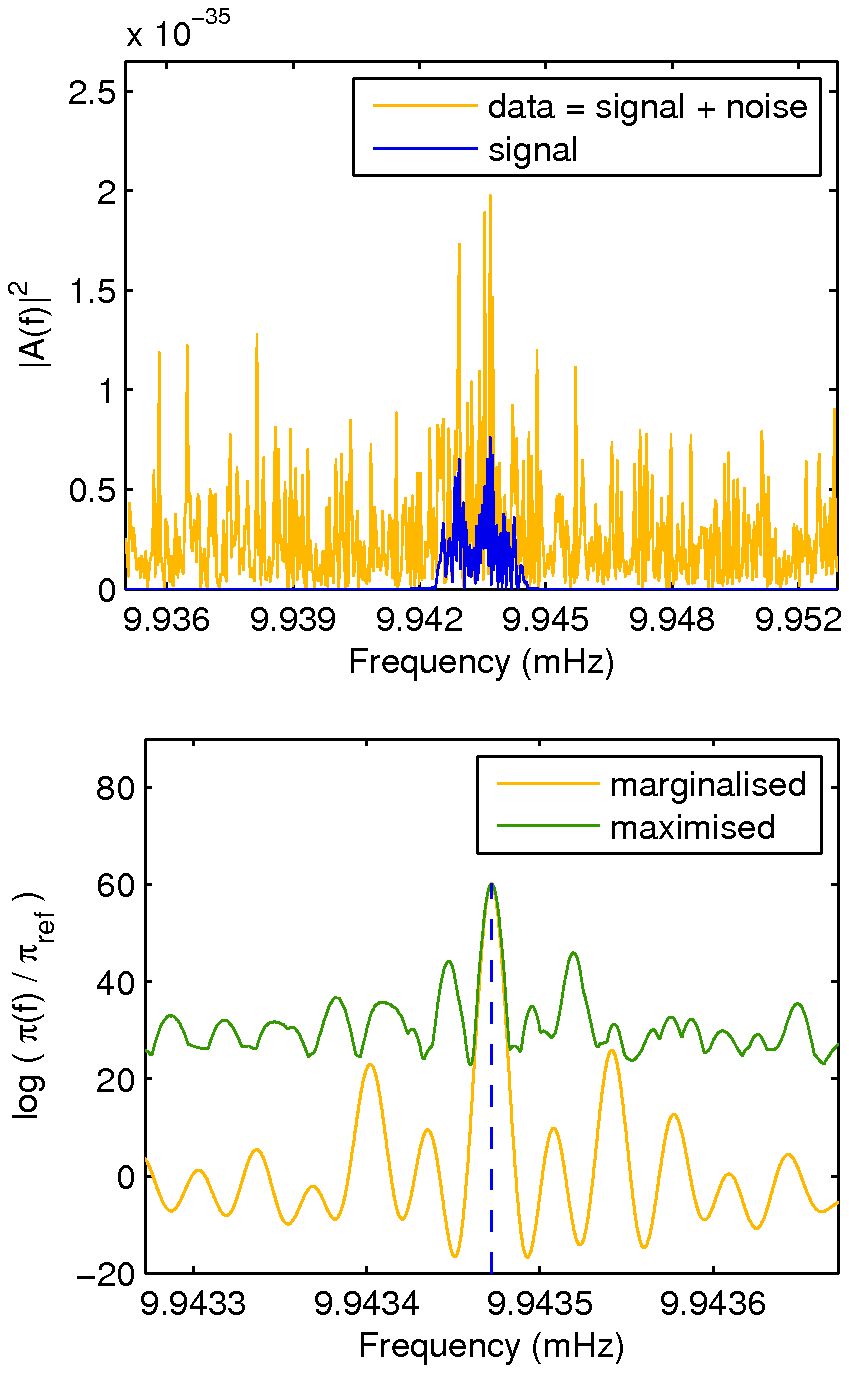} \\
(a) & (b)
\end{tabular}
\end{center}
\caption{Representative example of a (mock) LISA data set and the signal that we want to identify in the (a) time and (b) frequency domains. This particular signal corresponds to the (simulated) LISA's response to a gravitational wave emitted by a stellar-mass compact binary object generating a monochromatic signal at $\approx 10~\mHz$, that is modulated in both amplitude and phase, due to the change of position and orientation of the detector during the $1~\yr$ of observational time; the resulting signal-to-noise ratio is $15$. The top panel of (b) represents the Fourier transform of the data and signal sets plotted in (a), where we can see that the power is spread over several frequency bins because of the LISA motion, and the bottom plot of (b) corresponds to the marginalised and maximised likelihood function over all the parameters, $7$, except the frequency of the signal; the dashed line shows the actual value of the signal's frequency. The multimodal structure observed in the frequency space is also present in the two angles that define the sky location (\eg see Figure~\ref{Fig.likelihood}, although there, another set of parameter values were used), which dramatically reduces the efficiency of the MCMC algorithm unless a DR algorithm is implemented; see output in Figure~\ref{Fig.results_application_and_correlations}.}
\label{f:example_signal}
\end{figure}

Let us consider a gravitational wave signal produced by a short-period solar-mass compact binary system in our galaxy that is sufficiently far from coalescence that its frequency does not change over the observational time\footnote{In reality the frequency derivative, $\dot{f}$, is always different from zero; however as long as $\dot{f} \, T_\mathrm{obs}^2 \ll 1$, which is satisfied for the majority of close binary systems in the LISA band, the change of frequency can be neglected. It is trivial (although it may affects the computation time) to account for the change of frequency during the observation time, but for sake of simplicity this is neglected here.}. For a distant observer at rest (or moving with constant velocity with respect to the source), such signal would simply be a sinusoid, and the analysis of the data completely trivial. However, stellar-mass compact objects radiating in the LISA frequency window are long-lived, and during the observation time $T_\mathrm{obs}$ of several years, the instrument changes position and orientation, which in turns produces phase and amplitude modulations in the radiation recorded at the instrument output. Such modulations depend on the source frequency and sky location. The power of the signal is spread over several frequency bins: LISA's change of orientation causes a frequency shift of width $2/T_\mathrm{obs} \approx 6.5\times 10^{-8}~\Hz$ and the instrument motion around the Sun with velocity $v_\oplus \approx 30$ km $\mathrm{s}^{-1}$ and period 1 yr produces a Doppler modulation with typical width $ \approx (v_\oplus/c) f_0 \approx 10^{-7}\,(f_0/1\,\mathrm{mHz})~\Hz$.  We note that these frequency shifts are much smaller than the range over which the instrument noise varies, and we will assume that the noise contribution (which is actually unknown) has a constant variance $\sigma_n^2$. In summary the total number of parameters that describe the problem -- \ie the dimensionality of $\vec{\zeta}$ -- is $M = 8$:
\bea
\vec{\zeta}_\mathrm{s} & = & \{f_0, \A, \phi, \theta, \psi, \iota, \varphi_0\}\,,
\nonumber\\
\vec{\zeta}_\mathrm{n} & = & \sigma_n^2\,,
\eea
where $f_0$ is the constant frequency of the signal (with respect to the Solar System Barycentre), $\A$ is the amplitude, the longitude $\phi$ and co-latitude $\theta$ are the two angles that define the source's sky location, $\psi$ is the polarization angle of the gravitational wave; the inclination angle between the angular momentum of the system and an unitary vector parallel to the line of sight is defined by $\iota$, and $\varphi_0$ is a constant that fixes the initial phase of the signal. An example of a typical signal in frequency and time domain is shown in Figure~\ref{f:example_signal}; and an example of the resulting target distribution, that presents a multimodal structure, in particular in the frequency space and also in the two angles that define the sky location, is shown in Figure~\ref{Fig.likelihood}. Thus, for this particular problem it is crucial to have an MCMC algorithm capable of efficiently exploring the different maxima of the target distribution, because otherwise one can get in trouble trying to find the right signal inside the data as we show in \cite{TriasVecchioVeitch:2008}.

For the specific example presented here, we consider a mock LISA data set taken from the release 1B of the Mock LISA Data Challenges \citep{Babak-et-al:2008} and in particular, the data set MLDC-1B.1.1c  that contains a single source at frequency around $10~\mHz$ added to zero-mean Gaussian and stationary noise with signal-to-noise ratio of $15$. The implementation of the Delayed Rejection algorithm used in the analysis is done by following the steps discussed in Section~\ref{subsec:parameters}, \ie we start analysing the structure of the target distribution, and we derive that the optimal parameters for our proposals in frequency are $\sigma_1 = 0.45~\yr^{-1}$; $\sigma_2 = 0.2~\yr^{-1}$ and $\mu = 1.25~\yr^{-1}$. By looking at the appropriate plot of Figure~\ref{Fig.Losses_CPE} (in our case it would be approximately in between the second and third rows of third column) one realises that for the problem at hand $N_b = 0.95$ is an appropriate value, since the expected value of the ratio of proposals is $\sim e^{-2}$. The results are almost independent of $\nDR$, so the choice is driven mainly by computational reasons and we set  $\nDR = 2000$. Finally, the corresponding plot of Figure~\ref{Fig.Losses_AP} gives an idea of the $N_a$ value that one has to choose, low enough to mainly start the DR with a big jump in the frequency space, but not too much in order to avoid a negligible acceptance probability; in our case we'll work with $N_a = 0.15$. During the normal run, we enter into the DR routine with a probability of $10^{-3}$.

The correlations between the frequency, $f_0$, and the sky location, $\{ \phi \, , \, \theta \}$, force us also to update these two parameters during the DR chain, in order to prevent the efficiency to become very small. However, since the multimodal structure of the target distribution as a function of these two parameters is smoother, it is sufficient to draw their values from single Gaussian distributions, which reduces the irreversibility of the proposals to just the signal frequency, \ie all the results obtained for a single parameter in Sec.~\ref{sec:golden_rules} are completely valid here.

\begin{figure}
\begin{center}
\begin{tabular}{cc}
\includegraphics[width=0.445\textwidth]{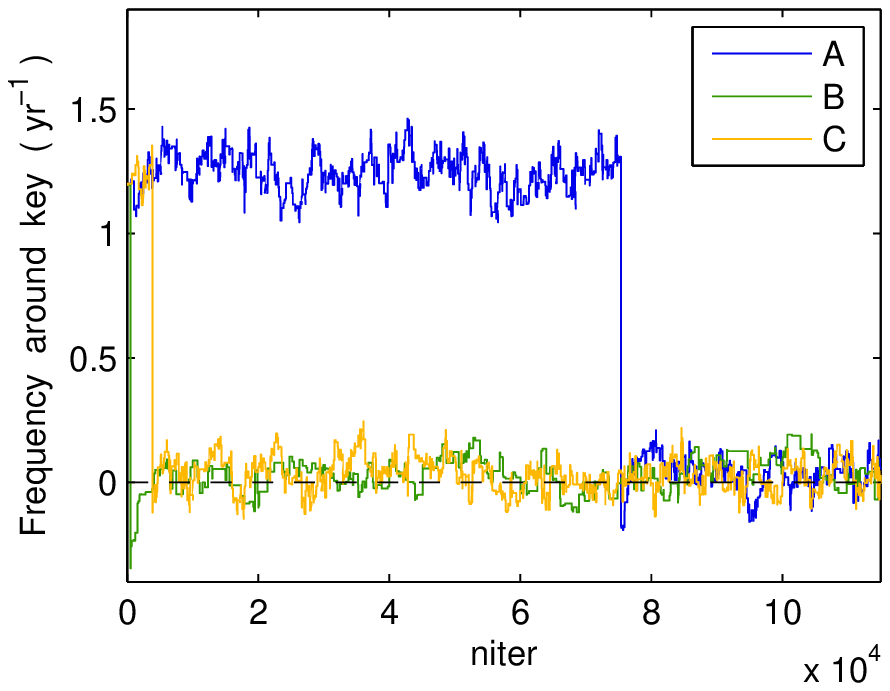} 		&
\includegraphics[width=0.445\textwidth]{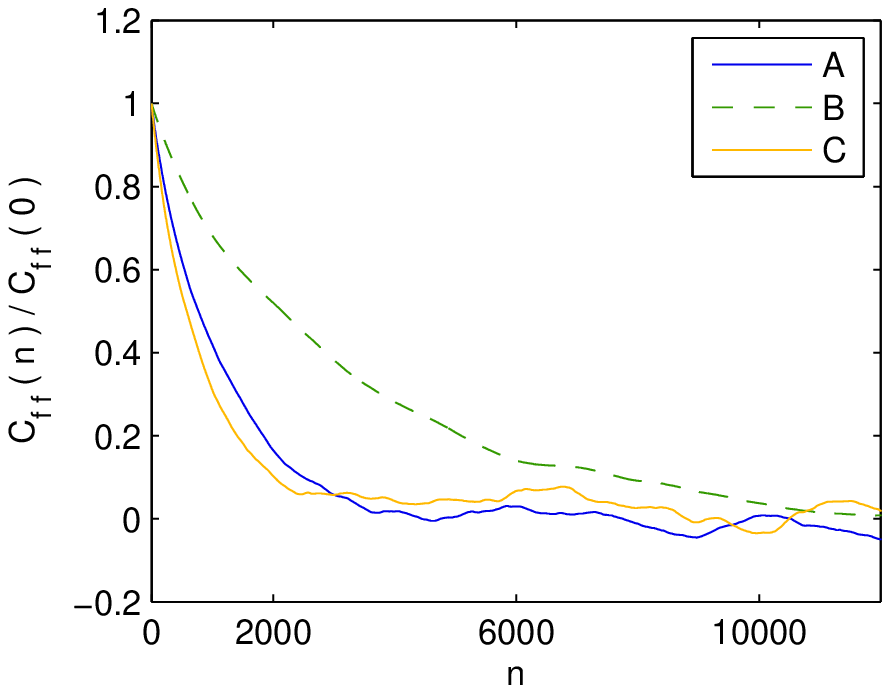} 		\\
\includegraphics[width=0.445\textwidth]{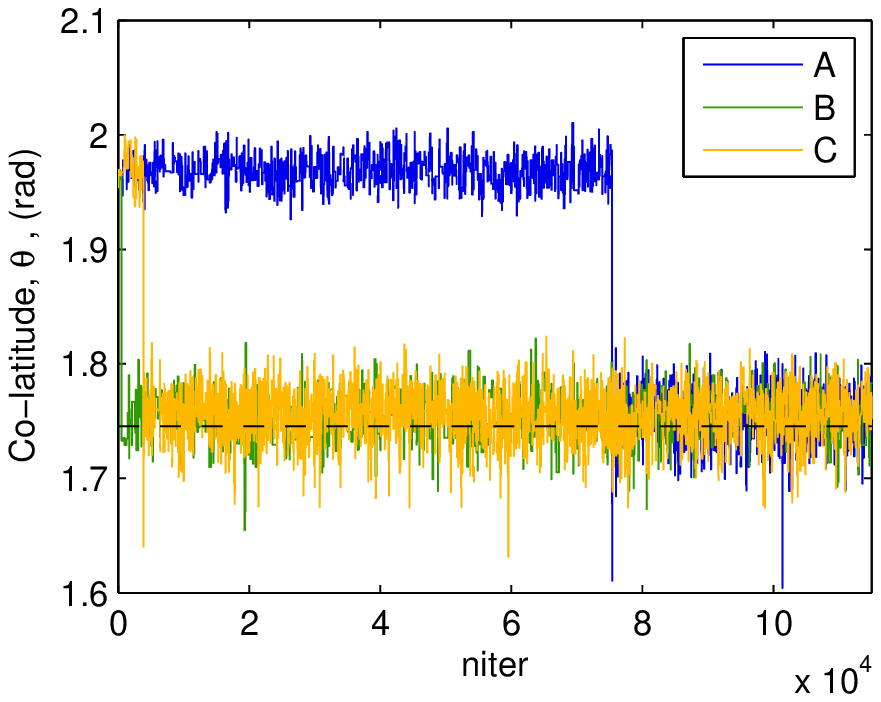} 		&
\includegraphics[width=0.445\textwidth]{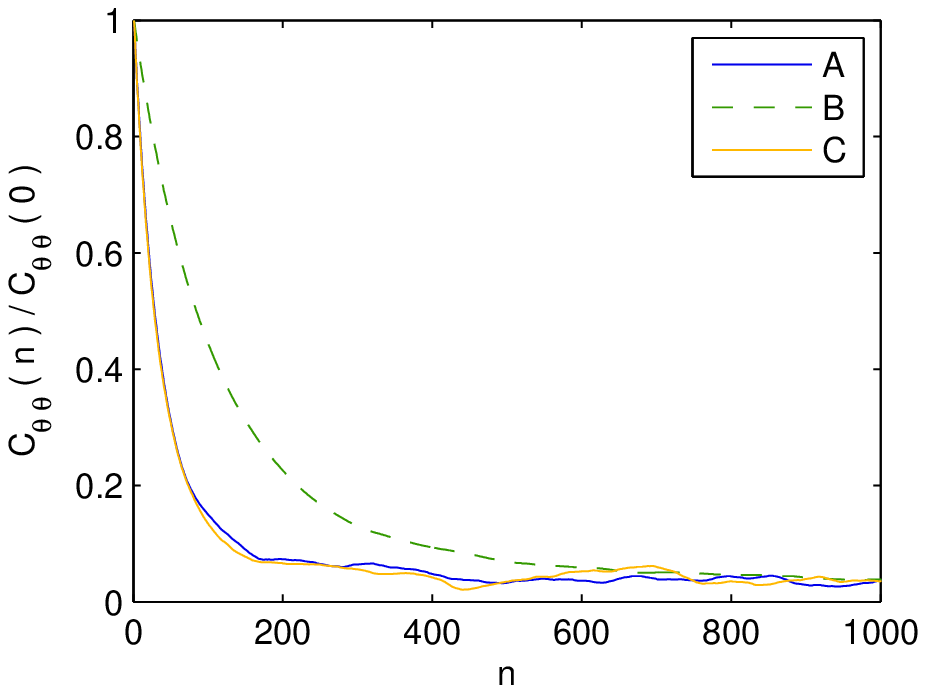} 		\\
\includegraphics[width=0.445\textwidth]{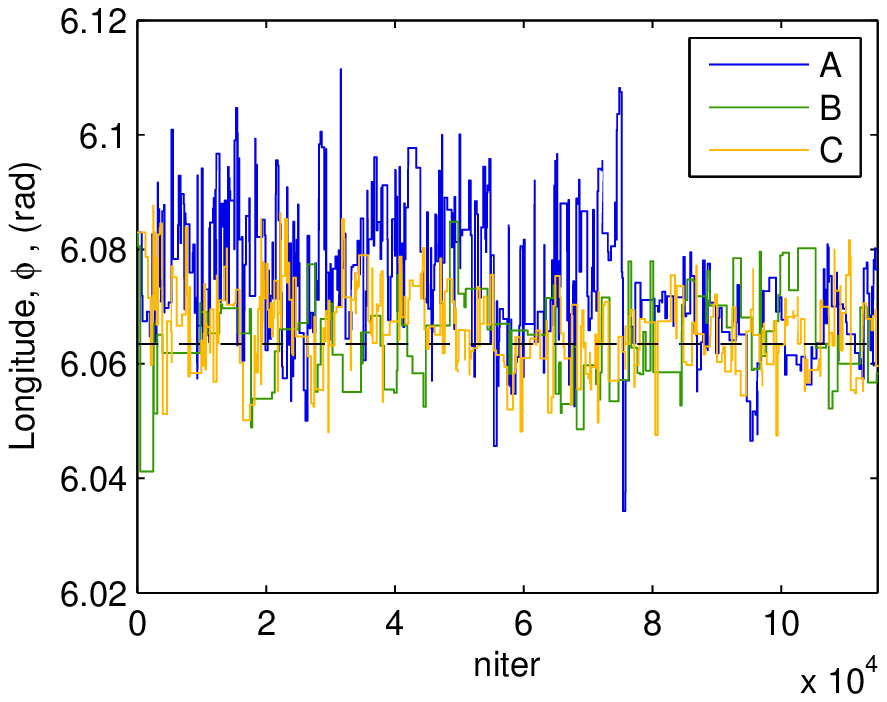}		&
\includegraphics[width=0.445\textwidth]{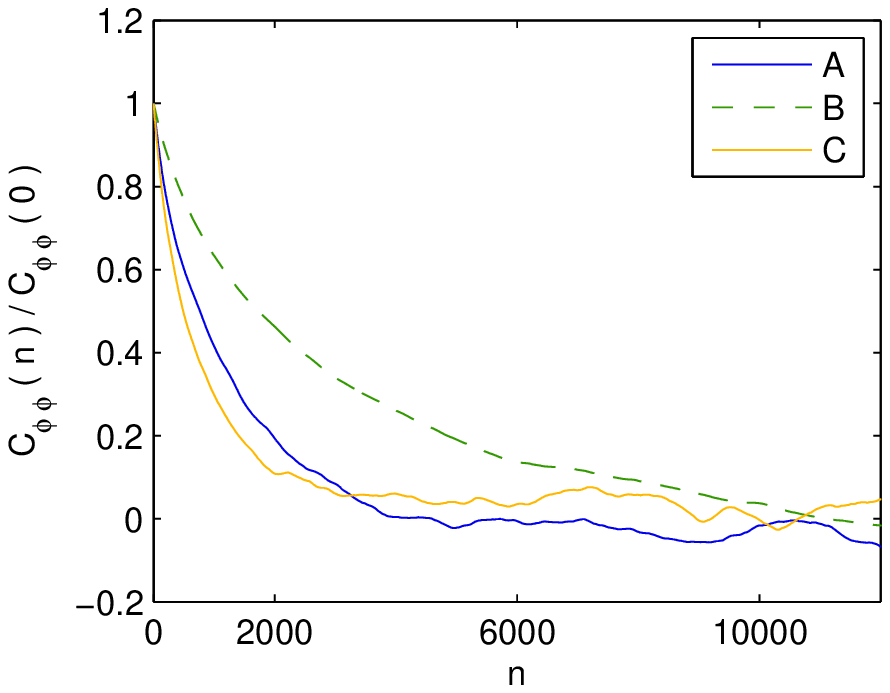}
\end{tabular}
\end{center}
\caption{Markov chains (left panels) and autocorrelation functions after making the transition to the main mode of the of the target distribution (right panels) obtained by searching for a gravitational wave signal from a single white dwarf binary observed by the Laser Interferometer Space Antenna with three different (Delayed Rejection) Markov chain Monte Carlo algorithms: (A) a standard MCMC method attempting big jumps in the parameter space with probability $\pBJ = 10^{-3}$, (B) a standard MCMC with $\pBJ = 2/3$; and (C) the DR MCMC algorithm described in this paper, entering into the Delayed Rejection routine with a probability of $10^{-3}$ and allowing up to $\nDR = 2000$ iterations to explore the new region of the parameter space. For this particular example, the proposal distributions during the DR are a mixture of three Gaussian functions for the signal frequency (parametrised by $\sigma_1 = 0.45~\yr^{-1}$;  $\sigma_2 = 0.2~\yr^{-1}$; $\mu = 1.25~\yr^{-1}$;  $N_a = 0.15$; $N_b = 0.95$ and $\nDR = 2000$) and two single Gaussians for the two sky locations, $\{ \phi \, , \, \theta \}$. From top to bottom the plots refer to frequency $f$, co-latitude, $\theta$, and longitude, $\phi$.}
\label{Fig.results_application_and_correlations}
\end{figure}

In Fig.~\ref{Fig.results_application_and_correlations} we compare the performance of the DR MCMC implementation just described above (case C in this figure), with two approaches based on the application of standard MCMC, one proposing big jumps with probability $\pBJ = 10^{-3}$ (case A) and the other attempting them with higher probability, $\pBJ = 2/3$ (case B). Notice that the choice of case A is based on keeping the same proportion of elements in the chain generated from small and big jumps in the parameter space as in case C, whereas in case B it is the ratio of attempted transitions that is kept constant from case C. In order to make a fair comparison, it is also important to notice that the computational power required to generate $N$ elements with the DR MCMC algorithm (case C) is almost three times higher than for the other two cases\footnote{With the DR MCMC approach, on average, every $1000$ iterations we will have $999$ elements drawn from a standard proposal and $1$ DR involving $\nDR = 2000$ evaluations of the likelihood function; so their computational cost (which is dominated by the likelihood evaluation that requires a time $t_{lik. eval.}$) will be $2999~t_{lik. eval.}$ in comparison to the $1000~t_{lik. eval.}$ of a standard MCMC.} (A and B), \ie given a certain computational time, we will be able to generate $3N$ elements of the chains for cases A and B and just $N$ elements for case C. For all these three cases, we assume the worst case scenario, starting the chains in a secondary maximum of the target distribution.

By looking at the chains (left panels of Fig.~\ref{Fig.results_application_and_correlations}) one see that either B or C are very efficient in exploring different maxima of the likelihood function, whereas the chain of case A needs $20$ times more iterations to accept such transition. On the other hand, when looking at the autocorrelation functions (right panels of Fig.~\ref{Fig.results_application_and_correlations}) we notice that the chain generated in case B is much more autocorrelated that the other two; in fact, one can get a more quantitative measurement of this quantity by computing the $\tau_{int}$ from its definition, Eq.~(\ref{eq:tau}): $(\tau_A , \tau_B , \tau_C )_{int, f} = (1112 , 3089 , 881)$; $(\tau_A , \tau_B , \tau_C )_{int, \theta} = (49.7 , 141.1 , 47.2)$ and $(\tau_A , \tau_B , \tau_C )_{int, \phi} = (1128 , 2813 , 868)$. Even taking into account that for the same computational time, $N_B \simeq 3 N_C$; the integrated autocorrelation times for B, $\tau_{int,B}$, are more than $3$ times larger than those of chain C, $\tau_{int, C}$ and therefore the variances obtained from C will be smaller than those from B.

In conclusion, we can see that the chain obtained from the DR approach (case C) is the preferred one when we combine the two requirements, having an algorithm that (i) efficiently explores multiple maxima and (ii) reduces the variance of the estimates once it has reached the stationary distribution.


\section{Conclusions and Future Work}
\label{sec:conclusion}

Multimodal target distributions characterise a wide variety of problems in many fields, reducing dramatically the efficiency of a Markov chain Monte Carlo sampling. Many of these problems have been successfully solved with the introduction of techniques that promote the movement of the chains by, \eg running parallel chains or making the target distribution temporarily smoother. These lines of attack fail however for complicated structures of the target distribution in many dimensions, that result in different localised `islands' where the function reaches high values (see Figure~\ref{Fig.likelihood}).

In this paper we have presented a fully Markovian algorithm capable of tackling these kind of problems when one has some previous knowledge about the target distribution structure, resulting in a efficient MCMC method to explore multimodal distributions. Our algorithm is based on the application of the Delayed Rejection scheme introduced by \cite{TierneyMira:1999, Mira:2001} with an arbitrary number of steps, whose successful implementation requires to tackle a number of non-trivial problems. In particular, we have explored the choice of proposal distributions to achieve finite acceptance probabilities (Sec.~\ref{sec:golden_rules}) and we have provided details for the numerical implementation of the algorithm (Sec.~\ref{sec:implementation_issues}). Finally, we have shown the benefits of this method using an example taken from the search for a gravitational wave signal in simulated data of the Laser Interferometer Space Antenna (Sec.~\ref{sec:application}).

An additional benefit of the Delayed Rejection algorithm described in this article is the possibility of combining the efficient exploration of the target distribution with the estimation of Bayes factors and performing model selection (for instance, using the Reversible Jump algorithm \citep{Green:1995}). This is a necessary extension of the approach presented here in order to successfully address the problem of searching for, and estimating the parameters of an unknown number of gravitational wave signals, which we are currently investigating.


\section*{Acknowledgments}

We would like to thank A.~Mira, C.~R\"{o}ver and R.~Umst\"{a}tter for helpful discussions about delayed rejection schemes and Markov chain Monte Carlo applications. The LISA's response function to gravitational wave signals from stellar binary systems used in Sec.~\ref{sec:application} is based on an implementation described in \cite{Cornish:2007if}. MT is supported by the Spanish Ministerio de Ciencia e Innovaci\'{o}n Research Projects FPA-2007-60220, HA2007-0042, CSD207-00042 and the Govern de les Illes Balears, Conselleria d'Economia, Hisenda i Innovaci\'{o}. AV and JV are supported by the UK Science and Technology Facilities Council.

\begin{appendix}

\section{Analytical expressions}
\label{ap:Analytical_results}

In Section~\ref{sec:golden_rules}, we have analysed the ratio of proposal distributions that appear in Equation~(\ref{eq:mastereq}), providing numerical results for its expected value in all the parameter space (Figs.~\ref{Fig.Losses_AP} and \ref{Fig.Losses_CPE}). In this section we try to do the same analytically, getting an expression for the losses due to make an asymmetrical proposal (AP), Figure~\ref{Fig.Losses_AP}, valid when the three Gaussians are well separated (\ie $\sigma_i / \mu$ small) and understanding why it's so important to always propose values from a distribution centred at the same location. We also justify why the other results can only be obtained numerically.

In general, using Bayes' theorem, one can derive an expression for the probability distribution of $f'=f(x)$, $x$ being a random variable generated from a known probability density function, $q(x)$:
\beq \label{eq:combined_prob}
p \left( f' = f(x) \right) = \sum_i \frac{ q(x_i)}{ |\frac{\partial f(x)}{\partial x} |_{x_i} } \qquad x_i \; | \; f(x_i) = f' \; .
\eeq
The derivation is made rewriting $p(f' = f(x))$ as the marginalised probability distribution of $p(f(x),x)$ over $x$, and then writing the joint probability function, $p(f(x),x)$, as the product of $p(x)$ times $p(f(x) | x)$, this last term being a delta function: $\delta(f'-f(x))$. After expanding the delta function as the sum over all $x_i$ such that $f(x_i) = f'$ one finally gets Eq.~(\ref{eq:combined_prob}).  

We want to apply the result from Eq.~(\ref{eq:combined_prob}) to compute the probability distribution of every factor that appear in Eq.~(\ref{eq:prop_ratio}), \ie the ratio of proposals from the final acceptance probability. In order to do this, we need to know the probability distribution from where $\betav_i$ were drawn (3-Gaussian functions, as is discussed in Sec.~\ref{subsec:3_gauss}) and the functions we use to evaluate the elements of the chain (also 3-Gaussian distributions, with not necessarily the same parameters). Next step would be finding all $x_i$ such that $f(x_i) = f'$, which for intermediate values of $f'$ and having 3-Gaussian functions would provide six solutions. In order to simplify the analytical calculations, we make use of the fact that the proposing and evaluating functions are always almost centred at the same location and we assume that their three modes are well separated ($\mu >> \sigma_i$), which allows to only have to evaluate in Eq.~(\ref{eq:combined_prob}) contributions from a single Gaussian for each $x_i$; \ie Equation~(\ref{eq:combined_prob}) will be a sum over six terms, where $i = \{1 , 2 \}$ will only involve contributions from the leftmost Gaussian, $i = \{3 , 4 \}$ from the central mode and $i = \{5 , 6 \}$ from the rightmost one.

We can distinguish three kind of terms from Equation~(\ref{eq:prop_ratio}): (a) all the terms in the denominator, which involve evaluating the elements of the chain with the same functions they were generated; (b) a couple of terms in the numerator in which the elements are evaluated with a 3-Gaussian function, with different relative weight of the modes as the proposal distribution used to generate them; and finally, (c) the rest of the terms, that evaluate the chain with a function a little bit shifted (\ie not centred at the same location) with respect to the actual proposal. In what follows, we are computing the probability distribution of each of these mentioned terms with the aim to combine the results and get an analytical expression for the probability distributions of the proposals ratio, Eq.~(\ref{eq:prop_ratio}):

\begin{itemize}
\item All the terms in the denominator, $q_j(\lambdav, \betav_1, \ldots, \betav_j)$, correspond to evaluate the elements of the chain with the same functions as they were generated. Assuming that both 3-Gaussian functions can be parametrised by\footnote{See Equation~(\ref{eq:Proposals}) and Figure~\ref{Fig.Proposals} for the meaning of each parameter.} $\{ \sigma_1 , \sigma_2 , \mu , N \}$ and making use of the result of Eq.~(\ref{eq:combined_prob}), one gets the expected distribution for a single term of the denominator of Equation~(\ref{eq:prop_ratio}), let us call it $f'$,
\beq \label{eq:prob_same_3gauss}
p(\log f') = \frac{2 \sigma_1 e^{\log f'}}{\sqrt{-2 \log(\sqrt{2\pi} \frac{\sigma_1}{N} f')}} + \frac{4 \sigma_2 e^{\log f'}}{\sqrt{-2 \log(\sqrt{2\pi} \frac{2 \sigma_2}{1-N} f')}}
\eeq
\beq \label{eq:mean_same_3gauss}
E \left[ \log f' \right] = -\frac{1}{2} - N \log\left(\sqrt{2\pi} \frac{\sigma_1}{N}\right) - (1-N) \log\left(\sqrt{2\pi} \frac{2 \sigma_2}{1-N}\right)
\eeq 

\item In the numerator, we evaluate the elements in reverse order, which produces terms with two kinds of asymmetries
\begin{itemize}
\item In the first and last elements of the product (see Eq.~(\ref{Eq.terms_AP})), we are evaluating a set of random variables generated from a 3-Gaussian distribution parametrised by $\{ \sigma_1 , \sigma_2 , \mu , N \}$ with another 3-Gaussian function, identical to the first one except for the relative weight of each mode, \ie parametrised by $\{ \sigma_1 , \sigma_2 , \mu , M \}$. The expected distribution of one of these terms that appear in the numerator of Eq.~(\ref{eq:prop_ratio}) can also be derived analytically,
\beq \label{eq:prob_diffN_3gauss}
p(\log f') = \frac{2 \sigma_1 \frac{N}{M} e^{\log f'}}{\sqrt{-2 \log(\sqrt{2\pi} \frac{\sigma_1}{M} f')}} + \frac{4 \sigma_2 \frac{1-N}{1-M} e^{\log f'}}{\sqrt{-2 \log(\sqrt{2\pi} \frac{2 \sigma_2}{1-M} f')}}
\eeq
\beq \label{eq:mean_diffN_3gauss}
E \left[ \log f' \right] = -\frac{1}{2} - N \log\left(\sqrt{2\pi} \frac{\sigma_1}{M}\right) - (1-N) \log\left(\sqrt{2\pi} \frac{2 \sigma_2}{1-M}\right)
\eeq 

\item All the other terms, Eq.~(\ref{eq:terms_to_CPE}), have the same relative weights between the Gaussians that generate and evaluate the random variables, but it can happen that the central location of the evaluation function is a little shifted with respect to the distribution used to generate the chain. Thus, we have that $q(x)$ is parametrised by $\{ \sigma_1 , \sigma_2 , \mu , N \}$, but the elements drawn from it are evaluated with a function $f(x)$ which parameters are $\{ \sigma_1 , \sigma_2 , \mu + \Delta , N \}$. Following the same procedure, the expected distribution for any of the terms that appear in the numerator of Eq.~(\ref{eq:prop_ratio}), is
\bea \label{eq:prob_delta_3gauss}
p(\log f') & = &
\frac{2 \sigma_1 e^{-\frac{\Delta^2}{2 \sigma_1^2}} e^{\log f'} \cosh\left( \frac{\Delta}{\sigma_1} \sqrt{-2 \log(\sqrt{2\pi} \frac{\sigma_1}{N} f')} \right) }{\sqrt{-2 \log(\sqrt{2\pi} \frac{\sigma_1}{N} f')}} 
+ \nonumber \\
 & & \frac{4 \sigma_2 e^{-\frac{\Delta^2}{2 \sigma_2^2}} e^{\log f'} \cosh\left( \frac{\Delta}{\sigma_2} \sqrt{-2 \log(\sqrt{2\pi} \frac{2 \sigma_2}{1-N} f')} \right) }{\sqrt{-2 \log(\sqrt{2\pi} \frac{2 \sigma_2}{1-N} f')}} 
\eea
\beq \label{eq:mean_delta_3gauss}
E \left[ \log f' \right] = -\frac{1}{2} - N \left[ \log\left(\sqrt{2\pi} \frac{\sigma_1}{N}\right) + \frac{\Delta^2}{2 \sigma_1^2} \right] - (1-N) \left[ \log\left(\sqrt{2\pi} \frac{2 \sigma_2}{1-N}\right) + \frac{\Delta^2}{2 \sigma_2^2} \right]
\eeq 

\end{itemize}

\end{itemize}

Since the mean of the sum of several random variables is the sum of their means, the expected value of the losses in acceptance probability due to an asymmetrical proposal (see Sec.~\ref{subsec:3_gauss}) may be computed as the sum of four means values from the previous results of Eqs.~(\ref{eq:mean_same_3gauss}) and (\ref{eq:mean_diffN_3gauss}). The result is Equation~(\ref{eq:mean_3gauss}), which, due to the assumptions made here, is only valid when the distance between Gaussians, $\mu$, is much bigger than their typical width, $\sigma_i$. In Figure~\ref{Fig.Validity_range_AP} it is plotted the validity range of the analytical result, which agrees with the numerical results of Figure~\ref{Fig.Losses_AP} as $\sigma_i / \mu$ is smaller.

The expected standard deviations of the full ratio of proposal functions, can not easily be derived analytically from Eqs.~(\ref{eq:prob_same_3gauss}) -- (\ref{eq:mean_delta_3gauss}) because of the correlations between variables. By combining the results from Eqs.~(\ref{eq:mean_same_3gauss}) and (\ref{eq:mean_delta_3gauss}) (see the output result in Equation~(\ref{eq:mean_centralproposal})) one gets the idea of the crucial importance of working with some proposals always centred at almost the same location (\ie $\Delta \simeq 0$), but it's not possible to analytically derive a general expression for the total losses due the central proposal evolution (CPE) because of two reasons, (a) we don't know the behaviour of $\Delta$ as we add more terms to compute $\bar{x}$ and (b) even knowing the expected value of $\Delta$ for each stage, we couldn't straightforwardly use this result since $\Delta$ doesn't appear linearly in the expression for the losses due to CPE.

\end{appendix}


\bibliographystyle{chicago_mod}
\bibliography{TVV-DRpaper}

\begin{thebibliography}{}

\bibitem[\protect\citeauthoryear{Al-Awadhi, Hurn and Jennison}{Al-Awadhi {\it et al.}}{2004}]{Al-Awadhi:2004}
Al-Awadhi, F., Hurn, M. and Jennison, C. (2004)
\newblock Improving the acceptance rate of reversible jump MCMC proposals.
\newblock {\em Statistics and Probability Letters} {\bf 69} 189-198

\bibitem[\protect\citeauthoryear{Babak {\it et al.}}{Babak {\it et al.}}{2008}]{Babak-et-al:2008}
  Babak, S. {\it et al.} (2008)
\newblock   The Mock LISA Data Challenges: from Challenge~1B to Challenge~3.
\newblock   {\em Class.\ Quant.\ Grav.\ } {\bf 25} 184026. 
\newblock See also http://astrogravs.nasa.gov/docs/mldc.

\bibitem[\protect\citeauthoryear{Babak, Gair and Porter}{Babak {\it et al.}}{2009}]{Babak:2009}
Babak, S., Gair, J.~R. and Porter, E.~K. (2009)
\newblock {\em An algorithm for detection of extreme mass ratio inspirals in LISA data.}
\newblock (Preprint: arXiv: 0902.4133)

\bibitem[\protect\citeauthoryear{Bender, Danzmann and the LISA Study Team}{Bender {\it et al.}}{1998}]{Bender:1998}
Bender, P. and Danzmann, K. and the LISA Study Team (1998)
\newblock {\em Laser Interferometer Space Antenna for the Detection of Gravitational Waves, Pre-Phase A Report.}
\newblock \textbf{MPQ 233} (Garching: Max-Planck-Institut f\"ur Quantenoptik)

\bibitem[\protect\citeauthoryear{Cornish}{Cornish}{2008}]{Cornish:2008}
Cornish, N.~J. (2008)
\newblock {\em Detection Strategies for Extreme Mass Ratio Inspirals.}
\newblock (Preprint: arXiv: 0804.3323)

\bibitem[\protect\citeauthoryear{Cornish and Littenberg}{Cornish and Littenberg}{2007}]{Cornish:2007if}
Cornish, N.~J. and Littenberg T.~B. (2007)
\newblock Tests of Bayesian Model Selection Techniques for Gravitational Wave Astronomy.
\newblock {\em Phys.\ Rev.\  D} {\bf 76} 083006

\bibitem[\protect\citeauthoryear{Cornish and Porter}{Cornish and Porter}{2007}]{Cornish:2007}
Cornish, N.~J. and Porter, E.~K. (2007)
\newblock The search for massive black hole binaries with LISA.
\newblock {\em Class.\ Quant.\ Grav.\ } {\bf 24} 5729-5755

\bibitem[\protect\citeauthoryear{Crowder and Cornish}{Crowder and Cornish}{2007}]{Crowder:2007}
Crowder, J. and Cornish, N.~J. (2007)
\newblock Solution to the galactic foreground problem for LISA.
\newblock {\em Phys.\ Rev.\  D} {\bf 75} 043008

\bibitem[\protect\citeauthoryear{Cutler}{Cutler}{1998}]{Cutler:1998}
  Cutler, C. (1998)
\newblock  Angular resolution of the LISA gravitational wave detector.
\newblock {\em Phys.\ Rev.\  D} {\bf 57} 7089  

\bibitem[\protect\citeauthoryear{Gair, Porter, Babak and Barack}{Gair {\it et al.}}{2008}]{Gair:2008}
Gair, J.~R. {\it et al.} (2008)
\newblock A constrained Metropolis-Hastings search for EMRIs in the Mock LISA Data Challenge~1B.
\newblock {\em Class.\ Quant.\ Grav.\ } {\bf 25} 184030

\bibitem[\protect\citeauthoryear{Gamerman}{Gamerman}{1997}]{Gamerman:1997}
Gamerman, D. (1997)
\newblock {\em Markov Chain Monte Carlo: Stochastic Simulation of Bayesian Inference.}
\newblock (London: Chapman \& Hall)

\bibitem[\protect\citeauthoryear{Gelman, Roberts and Gilks}{Gelman {\it et al.}}{1996}]{Gelman:1996}
Gelman, A. {\it et al.} (1996)
\newblock Efficient Metropolis Jumping Rules.
\newblock {\em Bayesian Statistics} {\bf 5} 599-607, Oxford, Clarendon Press

\bibitem[\protect\citeauthoryear{Geyer}{Geyer}{1991}]{Geyer:1991}
Geyer, C.~J. (1991)
\newblock Markov chain Monte Carlo maximum likelihood.
\newblock {\em Computing Science and Statistics: Proceedings of the 23rd Symposium on the Interface} 156-163

\bibitem[\protect\citeauthoryear{Geyer and Thompson}{Geyer and Thompson}{1995}]{Geyer:1995}
Geyer, C.~J. and Thompson, E.~A. (1995)
\newblock Annealing Markov Chain Monte Carlo with applications to Ancestral Inference.
\newblock {\em J.\ Am.\ Statist.\ Ass.\ } {\bf 90} 909-920

\bibitem[\protect\citeauthoryear{Gilks, Richardson and Spiegelhalter}{Gilks {\it et al.}}{1995}]{Gilks:1995}
Gilks, W., Richardson and S., Spiegelhalter, D. (1995)
\newblock {\em Markov Chain Monte Carlo in practice.}
\newblock (London: Chapman \& Hall)

\bibitem[\protect\citeauthoryear{Green}{Green}{1995}]{Green:1995}
Green, P.~J. (1995)
\newblock Reversible jump MCMC computation and Bayesian model determination.
\newblock {\em Biometrika} {\bf 82} 711-732

\bibitem[\protect\citeauthoryear{Green and Mira}{Green and Mira}{2001}]{GreenMira:2001}
Green, P.~J. and Mira, A. (2001)
\newblock Delayed rejection in reversible jump Metropolis-Hastings.
\newblock {\em Biometrika} {\bf 88} 1035-1053

\bibitem[\protect\citeauthoryear{Haario, Laine, Mira and Saksman}{Haario {\it et al.}}{2006}]{Haario:2006}
Haario, H., Laine, M., Mira, A. and Saksman, E. (2006)
\newblock DRAM: Efficient adaptive MCMC.
\newblock {\em Statistics and Computing}, vol. 16, num. 4, 339-354

\bibitem[\protect\citeauthoryear{Harkness and Green}{Harkness and Green}{2000}]{Harkness:2000}
Harkness, M.~A. and Green, P.~J. (2000)
\newblock Parallel chains, delayed rejection and reversible jump MCMC for object recognition.
\newblock {\em British Machine Vision Conference}

\bibitem[\protect\citeauthoryear{Hastings}{Hastings}{1970}]{Hastings:1970}
Hastings, W. (1970)
\newblock Monte Carlo sampling methods using Markov chains and their applications.
\newblock {\em Biometrika} {\bf 57} 97-109

\bibitem[\protect\citeauthoryear{Kirkpatrick, Gelatt and Vecch}{Kirkpatrick {\it et al.}}{1983}]{Kirkpatrick:1983}
Kirkpatrick, S., Gelatt, C.~D. and Vecchi, M.~P. (1983)
\newblock Optimization by Simulated Annealing.
\newblock {\em Science} {\bf 220} 671-680

\bibitem[\protect\citeauthoryear{Littenberg and Cornish}{Littenberg and Cornish}{2009}]{Littenberg:2009}
Littenberg, T.~B. and Cornish, N.~J. (2009)
\newblock {\em A Bayesian Approach to the Detection Problem in Gravitational Wave Astronomy.}
\newblock (Preprint: arXiv: 0902.0368)

\bibitem[\protect\citeauthoryear{Marinari and Parisi}{Marinari and Parisi}{1992}]{Marinari:1992}
Marinari, E. and Parisi, G. (1992)
\newblock Simulated Tempering: a New Monte Carlo Scheme.
\newblock {\em Europhysics Letters} {\bf 19} 451

\bibitem[\protect\citeauthoryear{Mira}{Mira}{2001}]{Mira:2001}
Mira, A. (2001)
\newblock On Metropolis-Hastings algorithms with delayed rejection.
\newblock {\em Metron} {\bf LIX} (3-4) 231

\bibitem[\protect\citeauthoryear{Mira}{Mira}{2002}]{Mira:2002}
Mira, A. (2002)
\newblock Ordering and improving the performance of Monte Carlo Markov Chains.
\newblock {\em Statistical Science} {\bf 16} 340

\bibitem[\protect\citeauthoryear{Peskun}{Peskun}{1973}]{Peskun:1973}
Peskun, P.~H. (1973)
\newblock Optimum Monte Carlo sampling using Makov chains.
\newblock {\em Biometrika} {\bf 60} 607

\bibitem[\protect\citeauthoryear{Raggi}{Raggi}{2005}]{Raggi:2005}
Raggi, D. (2005)
\newblock Adaptive MCMC methods for inference on affine stochastic volatility models with jumps.
\newblock {\em Econometrics Journal} {\bf 8} 235-250

\bibitem[\protect\citeauthoryear{Robert and Mengersen}{Robert and Mengersen}{2003}]{Robert}
Robert C.~P. and Mengersen, L. (2003)
\newblock IID sampling with self-avoiding particle filters: the pinball sampler.
\newblock {\em Bayesian Statistics} {\bf 7} 277-292, Oxford, Clarendon Press

\bibitem[\protect\citeauthoryear{Sokal}{Sokal}{1989}]{Sokal:1989}
Sokal, A.~D. (1989)
\newblock Monte Carlo methods in statistical mechanics: foundations and new algorithms.
\newblock {\em Cours de Troisi\`{e}me Cycle de la Physique en Suisse Romande}, Lausanne

\bibitem[\protect\citeauthoryear{Tierney and Mira}{Tierney and Mira}{1999}]{TierneyMira:1999}
Tierney, L. and Mira, A. (1999)
\newblock Some adaptive Monte Carlo methods for Bayesian inference.
\newblock {\em Statist.\ Med.\ } {\bf 18} 2507

\bibitem[\protect\citeauthoryear{Trias, Vecchio and Veitch}{Trias {\it et al.}}{2008}]{TriasVecchioVeitch:2008}
  Trias, M., Vecchio, A. and Veitch, J. (2008)
\newblock   Markov chain Monte Carlo searches for Galactic binaries in Mock LISA Data Challenge 1B data sets.
\newblock  {\em Class.\ Quant.\ Grav.\ }  {\bf 25} 184028 

\bibitem[\protect\citeauthoryear{Umst\"{a}tter {\it et al.}}{Umst\"{a}tter {\it et al.}}{2004}]{Umstatter:2004}
Umst\"{a}tter, R. {\it et al.} (2004)
\newblock Estimating the parameters of gravitational waves from neutron stars using an adaptive MCMC method.
\newblock {\em Class.\ Quant.\ Grav.\ } {\bf 21} S1655-S1665

\end{thebibliography}

\end{document}